%% file: colombia.tex
\documentclass[12pt]{article}
\usepackage{epsfig}
\usepackage{latexsym}
\usepackage{chart}
\newcommand{\psix}{\mbox{$\psi (x)$}}
\input{preamble}

\begin{document}
\baselineskip 24pt
\begin{center}
{\large\bf Concepts in Gauge Theory Leading to Electric--Magnetic
Duality}\\
\ \\
TSOU Sheung Tsun\\
Mathematical Institute, Oxford University\\
24--29 St.\ Giles', Oxford OX1 3LB\\
United Kingdom.\\
tsou\,@\,maths.ox.ac.uk
\end{center}

{\bf Abstract}\\
\baselineskip 14 pt

Gauge theory, which is the basis of all particle physics, is itself
based on a few fundamental concepts, the consequences of which are
often as beautiful as they are deep.  In this short lecture course I
shall try to give an introduction to these concepts, both from the
physical and mathematical points of view.  Then I shall show how these
considerations lead to a nonabelian generalization of the well-known
electric--magnetic duality in electromagnetism.  I shall end by
sketching some of the many consequences in quantum field theory 
that this duality engenders in particle physics.

These are notes from a lecture course given in the Summer School on 
{\em Geometric Methods in Quantum Field Theory}, Villa de Leyva, 
Colombia, July 1999, as well as a series of graduate lectures given 
in Oxford in Trinity Term of 1999 and 2000. 

\vspace*{5mm}
{\bf Synopsis}
\begin{itemize}
\item Gauge invariance, potentials, fields
\item Yang--Mills theory in action---the Standard Model of particle
physics
\item Principal bundles, connections, curvatures
\item Gauge group and charges
\item Action principle and symmetry breaking
\item Electric--magnetic duality
\end{itemize}

\clearpage

\section*{Introduction}
\hspace*{\parindent}In this lecture course I shall be mostly 
concerned about some basic
concepts of gauge theory, mainly to answer the question: why
are they introduced in physics?  Almost all of these concepts appear
already at the classical level, so that we shall deal principally with
the motion of a quantum charged particle in a classical (gauge)
field.  In so doing we shall find that certain fundamental questions
will arise which are usually not addressed by physicists, because they
are usually overwhelmed by many seemingly more pressing questions.

As a by-product, I shall also try to clarify a few notations to make
it easier for mathematicians to read physics literature.

Since Maxwell's theory of electromagnetism, that is, abelian gauge
theory, is the best understood gauge theory, I shall take it as our
reference point.  So most definitions will start with the abelian
case, and most results will be checked against it.  However, we shall
be extremely careful in {\em not assuming} abelian results for the
nonabelian case.

For most of the basic concepts, I shall be following my book.  
Apart from other textbooks both in mathematics and
physics which I have found useful, I have included in the
bibliography only a few papers.  For those who are interested to
pursue further, the references in the cited material will easily lead
them to other relevant articles.

\clearpage

\section{Gauge invariance, potentials, fields}
\hspace*{\parindent}In the first lecture, I shall use very simple 
mathematics only, so as
to put the emphasis on the {\em physical} reasons behind these
concepts.  In subsequent lectures, however, I shall not hesitate to
use more sophisticated mathematics, because it helps both in further
understanding and further developments.  In this way, I hope to show
you that it is the {\em physical situations} which force us to use
various sophisticated mathematical tools, and not the other way round, by
which I mean looking for or inventing physical theories to apply the
mathematical tools we have on hand.

\subsection{Notations and conventions}
\hspace*{\parindent}I shall use the following groups of terms synonymously:
\begin{itemize}
\item Maxwell theory, theory of electromagnetism, abelian theory;
\item Yang--Mills theory, nonabelian (gauge) theory (however, {\em
nonabelian} may take a truly mathematical meaning);
\item Spacetime, Minkowski space.
\end{itemize}

I shall use the following notations unless otherwise stated:
\begin{itemize}
\item $X$ = Minkowski space with signature $+---$
\item $\mu,\nu,\ldots$ = spacetime indices = 0,1,2,3
\item $i,j,\ldots$ = spatial indices or group indices
\item repeated indices are summed
\item $G$ = gauge group = compact, connected Lie group\\ (usually $U(n),
SU(n), O(n)$)
\end{itemize}

I shall make the following convenient assumptions: functions,
manifolds, etc. are as well behaved as necessary; typically functions
are continuous or smooth, manifolds are $C^\infty$.

I shall use the units conventional in particles physics, in which
$\hslash =1, c=1$, the former being the reduced Planck's constant and
the latter the speed of light.

{\underline{\em Caveat}}\ \  In a fully quantized field theory,
particles and fields
are synonymous.  Where there is no confusion, I shall use the two
terms interchangeably.

\subsection{Gauge invariance, potentials, fields: abelian case}
\hspace*{\parindent}Consider an electrically charged particle in an 
electromagnetic
field.  The wavefunction of this particle is a complex-valued function
$\psi (x)$ of $X$ (spacetime).  The phase of $\psi(x)$ is not a
measurable quantity, since only $|\psi (x)|^2$ can be measured and has
the meaning of the probability of finding the particle at $x$.  Hence
one is allowed to redefine the phase of $\psi (x)$ by an arbitrary
(continuous) rotation independently at every spacetime point without
altering the physics.  We say then that this theory possesses {\em
gauge invariance} or {\em gauge symmetry}.

In view of this arbitrariness, how can we compare the phases at
neighbouring points in spacetime?  In other words, how can we
`parallelly propagate' the phase?  We can, if we are given a
vector-valued function $A_\mu (x)$, called the {\em gauge potential}.
Then for the phases at $x$ and a nearby point $x+\Delta x$ to be
`parallel' we stipulate that the phase difference be $ e \amux \Delta
x^\mu$, where $e$ is a proportionality constant which will later be
identified with the charge of the particle.  This concept of
`parallellism' has to be consistent with gauge invariance.  In other
words, if we effect a rotation of $e \Lambda (x)$ on the phase of
\psix, 
$$
\psix \mapsto e^{ie \Lambda (x)} \psix
$$
then the phase rotation at $x+\Delta x^\mu$ will be
$$
e( \Lambda (x) + \partial_\mu \Lambda (x) \Delta x^\mu),
$$
so that for consistency the gauge potential must transform as
$$
\amux \mapsto \amux + \partial_\mu \Lambda.
$$

Next, by iterating the parallel phase transport along a given path 
$\Gamma$, we
can obtain a finite phase difference:
$$e \lineint{P}{Q}{\Gamma} A_\mu (x)\; dx^\mu. $$
This depends on the path in general.

\begin{figure}
\centering
\includegraphics{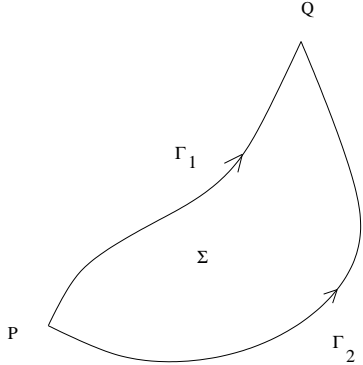}
\caption{Phase transport along a closed loop.}
\label{phasetransport}
\end{figure}
Now the phase difference (with $\Gamma=\Gamma_2 -\Gamma_1$)
$$ e \phi (\Gamma) = e \oint_\Gamma A_\mu dx^\mu = e \int_{\Gamma_2}
A_\mu dx^\mu - e \int_{\Gamma_1} A_\mu dx^\mu, $$
at the same point $P$ (see Figure \ref{phasetransport}) is a physically
measurable quantity (as observed
in interference phenomena), depends on the potential, and is in
general nonzero.  By Stokes' theorem
$$ e \phi(\Gamma) = -e \int\!\!\!\int_\Sigma \fdown (x) dx^\mu dx^\nu, $$
where $F_{\mu\nu} (x) = \partial_\nu A_\mu (x) - \partial_\mu A_\nu
(x). $  The quantity $-e F_{\mu\nu} (x) dx^\mu dx^\nu$ is the
infinitesimal phase change on going round the infinitesimal
parallelogram $ABCD$, Figure \ref{infinitransp}:
\begin{figure}[ht]
\begin{picture}(150,100)
\put(140,20){\framebox(80,50){}}
\put(130,0){$A$}
\put(220,0){$B$}
\put(130,80){$D$}
\put(220,80){$C$}
\put(150,10){\vector(1,0){15}}
\put(170,7){$\mu$}
\put(130,30){\vector(0,1){15}}
\put(128,50){$\nu$}
\put(240,20){$x+dx^\mu$}
\put(240,68){$x+dx^\mu+dx^\nu$}
\put(117,20){$x$}
\put(87,68){$x+dx^\nu$}
\end{picture}
\caption{Infinitesimal transport}
\label{infinitransp}
\end{figure}
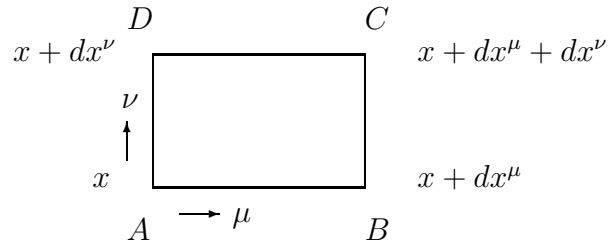

It follows that the tensor \fdown\ is a gauge invariant quantity, as
can be checked directly.  Now this change in phase, apart from the
factor $e$ representing the charge of the particle in question, acts
on all wavefunctions in a universal way and is hence a physical
property of the spacetime under consideration.  It represents
therefore a physical field, called the {\em electromagnetic field
tensor}. 

Historically, in classical physics, it was the components of this
field tensor, the electric and magnetic fields, which were introduced
first:
$$ F_{\mu\nu} = \left(
\begin{array}{cccc}
0 & E_1 & E_2 & E_3 \\
-E_1 & 0 & -B_3 & B_2 \\
-E_2 & B_3 & 0 & - B_1 \\
-E_3 & -B_2 & B_1 & 0
\end{array} \right).  $$
The potential was introduced as a mathematical convenience only, since
in classical physics it is not necessary to specify the potential,
only the field.

\subsubsection*{Bohm--Aharonov experiment}
{\hspace*{\parindent}}To demonstrate that in order to describe fully 
quantum mechanics in
electromagnetism one needs the potential, not just the field, Bohm and
Aharonov devised the following experiment, successfully performed by
Chambers. 
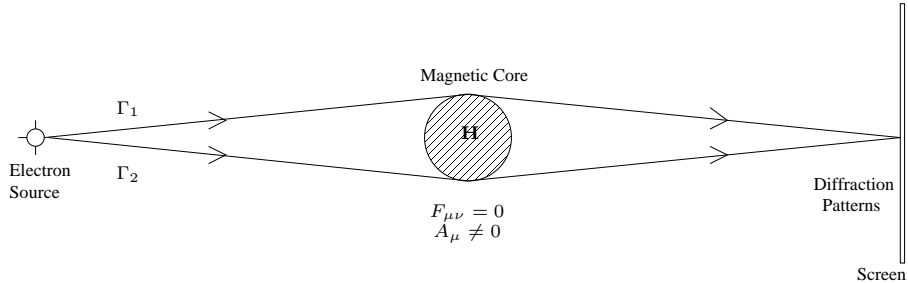
\begin{figure}[h]
\centering
\input{bohmaharo.pstex_t}
\caption{Schematic Representation of the Bohm-Aharonov Experiment.}
\label{Bohm}
\end{figure}
Electrons are made to travel along two different paths $\Gamma_1$ and
$\Gamma_2$, as illustrated in Figure \ref{Bohm}, enclosing a magnetic
core.  Outside this magnetic core, the field vanishes but the
potential is nonzero.  A diffraction pattern, made by the interference
of the two beams, is observed on the screen beyond the magnetic core,
and this diffraction pattern varies as the magnetic field and the
paths are varied, according
to the phase difference we calculated above.

This is one of the most important experiments of modern physics, and
its positive outcome is widely regarded as the strongest single
evidence supporting the basic tenets of electromagnetism as a gauge theory.

\subsection{Gauge invariance, potentials, fields: nonabelian case}
\hspace*{\parindent}Yang--Mills theory, as originally proposed by 
Yang and Mills in 1954,
is a generalization of electromagnetism in which the complex
wavefunction \psix\ of a charged particle is replaced by a
wavefunction with 2 components $\psi=\{ \psi^i (x),\ i=1,2\}$.  By a
change of `phase' we now mean a change in the orientation in {\em
internal space} of $\psi$ under the transformation:
$$\psi \mapsto S \psi,$$
where, to preserve probabilities, $S$ is to be a unitary matrix.  
In this way, {\em gauge invariance} can now be
interpreted as the requirement that physics be unchanged under
arbitrary $SU(2)$ transformations on $\psi$ independently (but
continuously) at different spactime points.  We note that it is
also possible to have the more general $U(2)$ transformations.

We can now consider parallel transport of this `nonabelian phase' in a
way similar to the abelian case.  Introducing the gauge potential
$A_\mu (x)$ as a matrix-valued spactime vector-valued function, we
again stipulate that at two neighbouring points $x$ and $x+\Delta
x$, the `phases' are parallel if they differ by $g A_\mu (x)
\Delta x^\mu$, where the porportionality contant $g$ again represents
a `charge':
$$\psi(x+\Delta x)=\exp (ig A_\mu (x)\Delta x^\mu)\,\psi (x). $$
Note that now $A_\mu (x) \in \mathfrak{su} (2)$, because it represents an
infinitesimal change in the phase of $\psi$.  Under a gauge
transformation $S(x) \colon \psi'(x) = S(x) \psi(x)$, so that 
$$ \psi'(x+\Delta x) = S (x+\Delta x) \psi (x+\Delta x).
$$
But
$$  \psi'(x+\Delta x) = \exp (ig A'_\mu (x) \Delta
x^\mu)\,\psi'(x), $$
where $A'_\mu$ is the gauge transform of $A_\mu$.

Equating, we get, for all \psix,
$$ S (x+\Delta x)\;\exp (ig A_\mu (x) \Delta x^\mu)\,\psi(x) =
\exp (ig A'_\mu (x) \Delta x^\mu)\,S(x)\,\psi (x).  $$
Expanding to leading order in $\Delta x$ and dropping \psix,
$$A'_\mu (x) = S(x)\,A_\mu (x)\,S^{-1} (x) - (\tfrac{i}{g})\, 
\partial_\mu S(x)\,S^{-1} (x). $$

When $S(x)$ is infinitesimal, in the sense
$$ S(x) = \exp (ig \Lambda (x)) \simeq 1 + ig\,\Lambda (x) $$
then we get
$$ A'_\mu (x)=A_\mu (x)+ \partial_\mu \Lambda (x) + ig[\Lambda(x),
A_\mu (x)], $$
which reduces to our previous formula in the abelian case. 

Next, we consider transport over a finite path $\Gamma$ as before.
Remembering that our primary concern is how the wavefunction $\psi$
changes, we see that what matters is the limit of the product of the
matrices $S$ as the step size tends to zero, and {\em not} the sum of
the `phases' $\Lambda$.
(Remember we are being unsophisticated!)  For matrices, in general $
e^{A+B} \ne e^A\,e^B$, so that the answer we are after is not
$$ \exp ig \int_\Gamma A_\mu (x) $$
but the product of $S$ as we said.  However, for sentimental reasons
about the abelian case, this product is usually written
$$\Phi (\Gamma) = P \exp ig\int_\Gamma A_\mu (x), $$
where the letter $P$ denotes
path-ordering. 
This Yang called the {\em Dirac phase factor}; it is also known as 
the {\em Wilson loop}. 

For a closed infinitesimal path in the form of a rectangle $ABCD$,
Figure \ref{infinitransp}, we can evaluate the change in phase as
before (where repeated indices on the right are not summed):
\begin{eqnarray*}
\psi_{ABC} (x+dx^\mu+dx^\nu)& =& \exp (ig A_\nu (x+dx^\mu)dx^\nu) \cdot
\exp (ig A_\mu (x)dx^\mu)\;\psi (x)\\
\psi_{ADC} (x+dx^\mu+dx^\nu)& =& \exp (ig A_\mu (x+dx^\nu)dx^\mu) \cdot
\exp (ig A_\nu (x)dx^\nu)\;\psi (x).
\end{eqnarray*}
Hence the phase difference at $A$, on expanding to leading order,
$$= \{ g(\partial_\mu A_\nu - \partial_\nu A_\mu) +ig^2 (A_\nu
A_\mu - A_\mu A_\nu)\}\;dx^\mu\,dx^\nu. $$
As before, we can define the field tensor $F_{\mu\nu} (x)$ by equating
this to $$-g\,F_{\mu\nu} (x)\,dx^\mu\,dx^\nu,$$ whence
$$F_{\mu\nu} (x)=\partial_\nu A_\mu (x) - \partial_\mu A_\nu (x) + ig
[A_\mu (x), A_\nu (x)]. $$

By similar considerations as before, we see that under a gauge
transformation $S(x)$, \fdown\ transforms as
$$F'_{\mu\nu} (x) = S(x)\,F_{\mu\nu} (x)\,S^{-1} (x). $$
We see that \fdown\ is no longer gauge invariant, as in the abelian
case, but gauge covariant.

\subsubsection*{Important Remarks}
\begin{enumerate}
\item In the above description of classical electromagnetism the gauge
group symmetry (that is, the phase freedom) plays no role (or only a
trivial one) if there are {\em no charges}, because $F_{\mu\nu}$, the
physical fields, are gauge invariant.  However, when we consider the
dynamics in the form of the Maxwell equations, then this symmetry
is the symmetry of those equations, and is therefore an important 
ingredient even in the classical field theory without charges.  This
remark does not apply to Yang--Mills theory, since there the field 
$F_{\mu\nu}$ being covariant and not invariant the gauge symmetry
should be taken into account already at the outset.
\item We note an extremely important difference betwen the abelian and
the nonabelian case: the finite phase difference is no longer related
to the surface integral of the field tensor in the nonabelian case.
First, the Dirac phase factor is not given by a line integral along
the closed path $\Gamma$.  In fact, the line integral has no
significance at all in the nonabelian case.  Moreover, the surface
integral of \fdown\ does not make sense, since there is no way to
order the matrices \fdown\ on a surface!
\item Because it is only {\em covariant} and not {\em invariant} the 
field tensor here is no longer a
physically measurable quantity, not even in classical physics.
\item Yang has shown that, both in abelian and nonabelian theories,
what describes the physics exactly is the Dirac phase factor
$\Phi(\Gamma)$ in the sense that the same physical situations
correspond to identical $\Phi(\Gamma)$ and different physical
situations to different $\Phi(\Gamma)$.  The obvious question then
arises: why do we not use $\Phi(\Gamma)$ as variables to describe
Yang--Mills theory and forget about the gauge dependent potential
\amux?  The answer is: the space of of closed loops is infinite
dimensional, which means that $\Phi(\Gamma)$ is more difficult to
handle, and there are vastly too many  $\Phi(\Gamma)$ variables.
Roughly speaking, we would be using functions of infinitely many
variables as opposed to 4 functions of 4 variables!  We shall have a
bit more to say on this later.
\end{enumerate}

\clearpage

\section{Yang--Mills theory in action---the Standard Model of particle
physics}
\hspace*{\parindent}Before we proceed any further and in order 
to exhibit the pivotal role
that Yang--Mills theory plays in modern physics, I have to tell you
some facts about particles.  This lecture is strictly for
mathematicians only!

\subsection{Particle classification}
\hspace*{\parindent}Particles are classified by their interactions 
among themselves and
with the fields in spacetime.  Apart from gravity, which we shall
neglect totally in these lectures, there are two kinds of
interactions: strong and electroweak.  Both are gauge theories, but
with a significant difference.  We shall take them in turn presently.

{\em Fundamental particles} are also distinguished by their spin.
Particles with integral spins (in suitable units) are called {\em
bosons}, and those with half-(odd)-integral spins are called 
{\em fermions}.  Known bosons all have spin 1, and are sometimes
called {\em vector bosons}.  There are some spin 0 bosons, called
{\em scalars}, which are postulated to exist but have not yet actually
been detected.  All known fermions have spin $\half$.  Fermions which
take part only in electroweak interactions are called {\em leptons}.

There are many (all unstable) particles called {\em resonances}
which are composites of those above.  (The only possibly stable
composite is the proton.)  They are mainly divided into
{\em mesons} (such as pions) or {\em baryons} (such as protons,
neutrons), but we shall not study them here.  There are more than 150
such so far discovered.

For our purposes, the following are the {\em fundamental} particles:

\smallskip
{\noindent}{\underline{Vector bosons}} (also known as gauge bosons): 
$\gamma$; $W^+,W^-,Z^0$; $g$

(photon; massive vector bosons; gluons)

\smallskip
{\noindent}{\underline{Leptons}}: $e,\nu_e$; $\mu, \nu_\mu$;
$\tau,\nu_\tau$

(electron, electron neutrino; muon, muon neutrino; tauon, tau
neutrino)

\smallskip
{\noindent}{\underline{Quarks}}: $u,d$; $c,s$; $t,b$

(up, down; charm, strange; top, bottom)

\smallskip
In a full quantum theory, these particles all have corresponding
antiparticles.  The notation is: the antiparticle of $x$ is denoted 
$\bar{x}$.

\subsubsection*{Remarks}

{\noindent}{\underline{Fact 1.}}  All known fundamental bosons have
spin 1.

\smallskip
{\noindent}{\underline{Fact 2.}}  All known fundamental fermions have
spin $\half$.

\smallskip
{\noindent}{\underline{Fancy 1.}} ($\stackrel{?}{\longrightarrow}$
{\underline{Fact 3.}})

Theory postulates existence of certain scalars called Higgs particles.

\smallskip
{\noindent}{\underline{Fancy 2.}} Supersymmetry needs spins $0, \half,
1, \tfrac{3}{2}, 2$.

\medskip
{\em Terminology changes as our understanding goes further.}

\subsection{The strong interaction}
\hspace*{\parindent}The strong interaction gives rise to nuclear 
forces and is governed by
an $SU(3)$ gauge theory.  This gauge symmetry is fancifully called
``colour'', and so the corresponding quantum field theory is usually
called {\em quantum chromodynamics} or QCD.

The gauge potential $A_\mu$ takes value in the Lie algebra
$\mathfrak{su} (3)$, and has hence 8 components.  Interpreting these
as particles (when the field is fully quantized), they form the
8-dimensional adjoint representation of $SU(3)$.  They are called the
{\em gluons}.  They are vector bosons and are massless.  They do not
have electric charges.

The massive particles are in the 3-dimensional fundamental
representation of $SU(3)$.  These are called {\em quarks}.  They are
spin $\half$ fermions, and have charges of $\tfrac{1}{3}$ or
$\tfrac{2}{3}$, in units of the electron charge.  There are 6 known
species of quarks: $u,d;\ c,s;\ t,b$; each of which has 3 components
corresponding to $SU(3)$.  These components are said to have different
``colours'', e.g.\ red, green and blue.  These have {\em no relation}
to the genuine colour that we see.

Just as the phase in electromagnetism can be arbitrarily rotated at
each spacetime point, and so cannot be measured, so is this colour
symmetry.  We say in both cases that the gauge symmetry is {\em
exact}.   Where QCD differs from QED is that particles with nontrivial
colour charges, i.e.\ in any representation of $SU(3)$ other than the
trivial one, cannot exist in the free state.   They cannot therefore
be directly observed, but indirectly their existence is fairly well
established experimentally.  This peculiar property goes under the
name of {\em confinement} and is special to {\em nonabelian}
theories.  To prove confinement is one of the most important aims of
theoretical physics at present.   Experimentally this is true so far,
since only singlets of $SU(3)$ have ever been observed.   This is the
probable origin of the name ``colour'', as in the hidden colours of
white light.

Quarks and antiquarks combine to produce observable particles as
resonances.  We can see which combinations are possible by looking at
the tensor product of the fundamental and conjugate fundamental
representations and picking out the singlet.  For example:
\begin{eqnarray*}
\mathbf{3} \otimes \bar{\mathbf{3}} & = & \mathbf{1} \oplus \cdots \\
\mathbf{3} \otimes \mathbf{3} \otimes \mathbf{3} & = & \mathbf{1}
\oplus \cdots 
\end{eqnarray*}
Thus we have: $$\pi^+=(u\bar{d}),\ \pi^-=(d\bar{u}),\
\pi^0=\tfrac{1}{\sqrt{2}} (u\bar{u}-d\bar{d})\,;\ p=(uud),\ n=(udd).$$
Ordinary mesons are $q\bar{q}$ states, and ordinary baryons are $qqq$
states.  In principle one can have higher composites and some such
exotics as $qq\bar{q}\bar{q}$ may have been already observed.

\subsection{The electroweak interaction}
\hspace*{\parindent}The electroweak interaction gives rise to both 
electromagnetic
phenomena and radioactivity, and is governed by a gauge theory with a
gauge group usually denoted as $SU(2) \times U(1)$.  (We shall give
more details about the specification of the exact gauge group in
Lecture 4.)  The $SU(2)$ part is often referred to as {\em (weak)
isospin}, and the $U(1)$ part as {\em (weak) hypercharge}.

In electroweak (or Weinberg--Salam) theory an important novel 
ingredient is introduced in
Yang--Mills theory, that is, {\em (spontaneous) symmetry breaking}.
In addition to the gauge bosons (vector bosons) and the massive
fermions as in QCD, we introduce some scalar (i.e.\ spin 0) particles
called {\em Higgs fields} $\phi$.   They are in a 2-dimensional
representation so that they are in fact gauge spinors:
$$\phi= \left( 
\begin{array}{c}
\phi^+ \\ \phi^0 
\end{array} \right). $$ 
The lowest energy state (called {\em vacuum}) occurs when $\phi \ne
0$, so that a physical system in such a vacuum state corresponding to
$\phi_0$ will no longer be invariant under the whole of $SU(2) \times
U(1)$, only under a $U(1)$ subgroup which leaves the spinor $\phi_0$
invariant.  This is the situation of symmetry breaking: although the
theory has $SU(2) \times U(1)$ invariance, the actual physical system
has a smaller invariance.  This $U(1)$ subgroup is generated by a
linear combination of $T_3$ and $Y$, where $T_1,T_2,T_3$ are the
generators of $\mathfrak{su} (2)$ and $Y$ is the generator of weak
hypercharge $U(1)$.  (We shall go into more details later.)

\medskip

{\noindent}{\bf Reminder.}\ \ 
The notation for $T_i$ is exactly the same as for ordinary spin, where
$T_3$ is represented by the diagonal matrix
$$T_3=- \frac{1}{2} \left(
\begin{array}{rr}
i & 0\\ 0 & -i 
\end{array} \right) $$
and the commutation relation is $[T_i,T_j]= \epsilon_{ijk} T_k$.

\medskip

The residual gauge group $U(1)$ is identified with the $U(1)$ gauge
group of electromagnetism, as observed in the physical world.

The mechanism of symmetry breaking is often visualized as a phase
change.  In the early universe when the temperature was high, the
whole $SU(2) \times U(1)$ symmetry was exact.  As the temperature
decreased to a critical value, symmetry breaking occurred and the
electromagnetic gauge symmetry ``froze out'' to produce the phase we
are in today.

As a result of the symmetry breaking, 3 of the 4 gauge bosons combine
with some components of the Higgs doublet $\phi$ to become massive
vector bosons ($W^+,W^-,Z^0$).  The boson corresponding to $U(1)_{\rm
em}$ remains massless and is the {\em photon} of the electromagnetic
field.

The leptons are the fermions of the theory.   The charged leptons are
$e,\mu,\tau$, and the neutral leptons are the {\em neutrinos}
$\nu_e,\nu_\mu,\nu_\tau$.   Each lepton wavefunction $\psi$ can be
projected into 2 mutually orthogonal components:
$$\psi=\psi_L +\psi_R.$$
Since $\psi$ has spin $\half$, it is a spacetime spinor and $\psi_L$
and $\psi_R$ are eigenstates of the {\em chirality operator}
$\gamma^5$.   Then the representation assignments are:

\medskip

$$\begin{array}{cccl}
\left( \begin{array}{c} \nu_e \\ e \end{array} \right)_L & 
\left( \begin{array}{c} \nu_\mu \\ \mu \end{array} \right)_L &
\left( \begin{array}{c} \nu_\tau \\ \tau \end{array} \right)_L &
SU(2)\ {\rm doublets} \\
\ \\
e_R & \mu_R & \tau_R & SU(2)\ {\rm singlets}
\end{array} $$

\medskip

Notice that neutrinos are supposed to have only left-handed components
in this assignment.  In view of the recent SuperKamiokande results on
neutrino oscillations, it may be necessary to revise this assignment
and suppose the neutrinos have also right-handed components (and a
nonzero mass).

\subsection{The Standard Model}
\hspace*{\parindent}The particle spectrum is summarized in 
Table \ref{smpart}. 
Note that the particles in brackets are not (or have not been)
directly observed, but they are part of the gauge theory.
\begin{table}
$$\begin{array}{|l|c|c|c|}  \hline
{\rm Force} & {\rm Gauge\ symmetry} & {\rm Gauge\ bosons} & 
{\rm Matter} \\ \hline\hline
{\rm Strong} & SU(3) & {\rm (gluons)} & {\rm (quarks)} \\
{\rm (QCD)} &&& \\ \hline
{\rm Electroweak} & SU(2) \times U(1) & \gamma; W^+,W^-,Z^0 & 
{\rm leptons} \\
{\rm (Weinberg-Salam)} &&& {\rm (Higgs)} \\ \hline
\end{array}$$
\caption{Particle content of the Standard Model}
\label{smpart}
\end{table}

The standard model is an amalgamation, a knitting together, of the
above two theories in such a way that {\em all} of known particle
physics, up to the present day, is encompassed.  It is a Yang--Mills
theory with a gauge group which is usually written as $SU(3) \times
SU(2) \times U(1)$---we shall examine this in more detail in Lecture
4.  The particle content can be schematically represented as 
\begin{center}
(QCD + Weinberg--Salam) $\times$ 3.
\end{center}
The multiplication by 3 is necessary to model another aspect of the
particle spectrum known as {\em generation}.  Take the charged leptons
as an example.  There are 3 of them: $e,\mu,\tau$.   Except for their
very different masses: 
$$m_\tau \colon m_\mu \colon m_e \cong 3000
\colon 200 \colon 1$$ 
they behave in extemely similar fashions.  The 3
neutrinos $\nu_e,\nu_\mu,\nu_\tau$ also have similar interactions.
The quarks also come in 3 generations: $u,c,t$ with charge
$\tfrac{2}{3}$ (called the {\em up-type quarks}) and $d,s,b$ with charge
$-\third$ (called the {\em down-type quarks}).

But the standard model is not just putting the 2 theories together,
because although the leptons do not transform under $SU(3)$ (since
they have no strong interaction), the quarks are in nontrivial
representations of weak isospin $SU(2)$.  In fact, we can set up Table
\ref{generations} for the 3 generation, and Table \ref{lepq} for the
lightest generation (and similar ones for the other two generations).
\begin{table}
$$\begin{array}{cc}
{\begin{array}{c} \uparrow \\
{\rm g}\\{\rm e}\\{\rm n}\\{\rm e}\\{\rm r}\\{\rm a}\\{\rm t}\\{\rm
i}\\{\rm o}\\{\rm n}\\{\rm s} \\ \downarrow \end{array}} &
{\begin{array}{|l|l|} \hline
{\rm Quarks} & {\rm Leptons} \\ \hline\hline
& \\
\left( \begin{array}{c} 
u \\ d \end{array} \right)_L, u_R, d_R & \left( \begin{array}{c}
\nu_e \\ e \end{array} \right)_L, e_R \\ 
& \\ \hline
& \\ 
\left( \begin{array}{c} 
c \\ s \end{array} \right)_L, c_R, s_R & \left( \begin{array}{c}
\nu_\mu \\ \mu \end{array} \right)_L, \mu_R \\ 
& \\ \hline
& \\
\left( \begin{array}{c}
t \\ b \end{array} \right)_L, t_R, b_R & \left( \begin{array}{c}
\nu_\tau \\ \tau \end{array} \right)_L, \tau_R \\ 
& \\
\hline
\end{array}} \end{array}$$
\caption{The 3 generations of leptons}
\label{generations}
\end{table}

\begin{table}
$$\begin{array}{|c||c|l|c|c|} \hline
& SU(3) & SU(2);\;I_3 & U(1)_Y & U(1)_{\rm em}  \\ \hline\hline
u_l & \mathbf{3} & \mathbf{2};\;\half & \third & \frac{2}{3} \\ \hline
d_l & \mathbf{3} & \mathbf{2};\;-\half & \third & \third \\ \hline
u_R & \mathbf{3} & \mathbf{1};\;0 & \tfrac{4}{3} & \frac{2}{3} \\
\hline
d_R & \mathbf{3} & \mathbf{1};\;0 & -\tfrac{2}{3} & \third \\
\hline\hline
\nu_{eL} &\mathbf{1} & \mathbf{2};\;\half & -1 &0 \\ \hline
e_L & \mathbf{1} &\mathbf{2};\;-\half & -1 & -1 \\ \hline
e_R & \mathbf{1} & \mathbf{1};\;0 & -2 & -1 \\ \hline
\end{array} $$
\caption{Charges of leptons}
\label{lepq}
\end{table}
In Table \ref{lepq} the electric charge satisfies: $Q=I_3 + \half Y$.
Other normalizations exist in the literature.

The standard model is a very good representation of the state of our
present knowledge, but most physicists doubt that it constitutes a
full final theory.

\subsubsection*{Good points}
\begin{enumerate}
\item Purports to explain {\em all} of particle physics.
\item Survives {\em all} precision tests so far.
\end{enumerate}

\subsubsection*{Open questions}
\begin{enumerate}
\item Why 3 copies (generations)?
\item Where do the Higgs fields come from?
\item Why are quarks confined?
\item There are over 20 parameters which are not explained by the theory.
\end{enumerate}

\clearpage

\section{Principal bundles, connections, curvatures}
\hspace*{\parindent}In this lecture we change tack altogether and 
make contact with
differential geometry.  Most mathematicians working on Yang--Mills
theory work on vector bundles, but following Yang I tend to think in
terms of principal bundles.  The two ways are of course equivalent.

In order to simplify definitions and so on and avoid all unnecessary
troubles, I shall make the following general assumption:
\begin{center}
`Things are as nice as possible.'
\end{center}
For example, manifolds and maps are smooth, Lie groups are compact
connected, equivalence classes can be confused with their
representatives.

Also, no formal proofs will be given.

\subsection{Principal bundles}
{\noindent}{\bf Definition.}\ \ {\it A {\em principal coordinate
bundle} $\mathcal{P}$ is a collection of the following:
\begin{enumerate}
\item a manifold $P$ called the {\em total space},
\item a manifold $X$ called the {\em base space},
\item a projection $\pi \colon P \to X$, with $\pi^{-1} (x), x \in X$,
called the {\em fibre above} $x$,
\item a Lie group $G$ acting on itself by left translation, called the
{\em structure group},
\item an open cover $\{U_\alpha\}_{\alpha \in \Lambda}$ of $X$,
\item $\forall \alpha \in \Lambda$, a diffeomorphism called {\em
coordinate function}
$$ \phi_\alpha \colon U_\alpha \times G \longrightarrow \pi^{-1}
(U_\alpha) $$
satisfying
\begin{enumerate}
\item $\pi \phi_\alpha (x,g)=x,\ \forall x \in U_\alpha, g \in G$,
\item if we define $\forall x \in U_\alpha$
\begin{eqnarray*}
\phi_{\alpha,x} \colon G & \to & \pi^{-1} (x) \\
g & \mapsto & \phi_\alpha (x,g)
\end{eqnarray*}
then on the overlap $U_\alpha \cap U_\beta$ the composite
$$ \phi^{-1}_{\beta,x} \phi_{\alpha,x} \colon G \to G $$
is the left multiplication by a uniquely determined element
$\phi_{\beta\alpha}$ of $G$,
\item the map
$$ \phi_{\beta\alpha} \colon U_\alpha \cap U_\beta \to G $$
is smooth---it is called the {\em transition} or {\em patching function}.
\end{enumerate}
\end{enumerate}}

\bigskip

\begin{figure}
\centering
\includegraphics{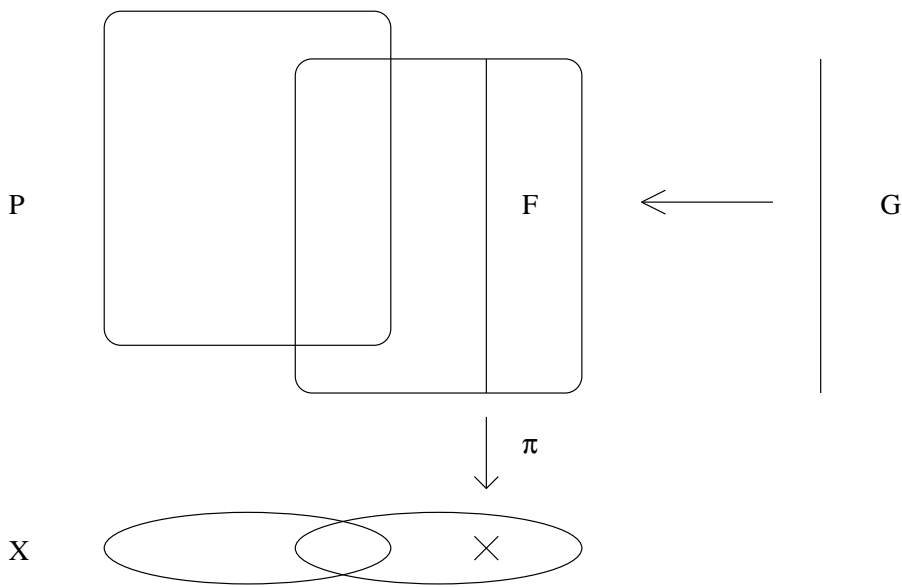}
\caption{Sketch of a principal bundle.}
\label{pbundle}
\end{figure}

A sketch (Figure \ref{pbundle}) may be helpful.

\bigskip

{\noindent}{\bf Definition.}\ \ {\it Two principal coordinate bundles
$\mathcal{P}$ and $\mathcal{P'}$ are said to be {\em strictly
equivalent} if they have the same total space $P$, the same base space
$X$, the same projection $\pi$, the same structure group $G$, and
their coordinate functions $\{\phi_\alpha\},\ \{\phi'_\beta\}$ are
such that the composite 
$$ \bar{\phi}_{\beta\alpha} (x) = \phi'^{-1}_{\beta,x}
\phi_{\alpha,x}, \quad x \in U_\alpha \cap U'_\beta $$
is left multiplication by an element of $G$.}

\bigskip

{\noindent}{\bf Definition.}\ \ {\it A {\em principal bundle} is an
equivalence class of coordinate principal bundles under strict
equivalence.}

\bigskip

{\noindent}{\bf Definition.}\ \ {\it A {\em trivial principal bundle}
is one in which 
$$ P \cong X \times G. $$}
This is the case in most applications to physics.

\bigskip

{\noindent}{\bf Note.}\ \ We shall often call $P$ the principal bundle.

\bigskip

{\noindent}\framebox{Dictionary 1}

\smallskip

\begin{tabular}{rcl}
base space & $\longleftrightarrow$ & spacetime\\
structure group & $\longleftrightarrow$ & gauge group\\
principal bundle & $\longleftrightarrow$ & gauge theory\\
principal coordinate bundle & $\longleftrightarrow$ & gauge theory in a
particular gauge
\end{tabular}

\subsection{Connections and curvatures}
\hspace*{\parindent}First we recall a few definitions.
\begin{enumerate}
\item Give a map $f \colon X \to X'$ we can define its {\em
differential} $f_*$ at $x \in X$, as a linear map $T_x X \to T_{f(x)}
X'$ as follows.  Given a tangent vector $U$ at $x$, choose any curve
$x(t)$ in $X$ such that $x(0) =x$ and $U$ is the tangent to $x(t)$ at
$x$.  Then the image $f_* U$ is the tangent vector to the image curve
in $X'$ at $f(x)$.  It can be shown that the definition is independent
of the curve $x(t)$.  Similarly, given any 1-form $\omega'$ on $X'$,
we can define a 1-form $f^* \omega'$ on $X$ by
$$ (f^* \omega') V = \omega' (f_* V), $$
for any  vector field $V$ on $X$.
\item Denote by $L_a$ the left multiplication by an element $a \in
G$.  Let $\mathfrak{g}$ be the Lie algebra of $G$.  A vector field $A$
on $G$ is said to be {\em left invariant} if $(L_a)_* A =A,\ \forall a
\in G$.  Recall then that $\mathfrak{g}$ can be characterized as the
set of left invariant vector fields on $G$.
\item Suppose a group $G$ acts on the right on a manifold $P$.  Then
for each $A \in \mathfrak{g}$, the action induces a vector field
$\sigma (A)$ on $P$ as follows.  At each $u \in P$, consider the
action of the 1-parameter subgroup $\exp tA$, whose orbit is a curve
in $P$ passing through $u$ at $t=0$.  The tangent to the curve at $u$
is the required vector.  We call $\sigma(A)$ the {\em fundamental
vector field} corresponding to $A \in \mathfrak{g}$.
\end{enumerate}

Now we come to the definition of a connection on $P$.  Consider the
action of $G$ on $P$ given by, $\forall a \in G,\ u \in P$,
$$ R_a (u)= \phi_\alpha (x, (\phi^{-1}_{\alpha,x} (u))a), $$
where $x=\pi (u) \in U_\alpha$.  Note that this action moves points
along the same fibre, and is indeed a right action since $R_{a_1 a_2}
(u)= R_{a_2} (R_{a_1} (u)),\ u \in P$.  We also write: $R_a (u) = ua$.

\medskip

{\noindent}{\bf Definition.}\ \ {\it Given a principal bundle $P$ as
above, a {\em connection 1-form} $\omega$ on $P$ is a
$\mathfrak{g}$-valued 1-form on $P$ satisfying
\begin{enumerate}
\item $\omega (\sigma (A)) =A,\ \forall A \in \mathfrak{g}$,
\item $\omega ((R_a)_* V) = {\rm ad} (a^{-1}) \omega (V),\ \forall a
\in G, \forall$ vector fields $V$ on $P$, where the adjoint action
ad$(a^{-1})$ of $a^{-1}$ on $A \in \mathfrak{g}$ is often written as
$a^{-1} A a$.
\end{enumerate}}

\medskip

We wish now to show how to define, using a set of local sections
$\{U_\alpha\}$, local 1-forms $\{ \omega_\alpha\}$ which are
equivalent to the given connection $\omega$. 

Choose local sections $u_\alpha \colon U_\alpha \to P$ such that
$$ \phi_\alpha (x,e) = u_\alpha (x), \quad x \in U_\alpha \cap
U_\beta. $$
Then $u_\beta (x) = u_\alpha (x) \phi_{\beta\alpha} (x). $  Let $V$ be
a tangent vector of $X$ at $x$; then
$$du_\alpha \colon T_x X \to T_{u(x)} P. $$
We have
\begin{eqnarray*}
du_\beta (V) & = & du_\alpha (V) \phi_{\beta\alpha} (x) + u_\alpha
d\phi_{\beta\alpha} (V) \\
& = & du_\alpha (V) \phi_{\beta\alpha} (x) + u_\alpha
\phi_{\beta\alpha} (x) (\phi_{\beta\alpha} (x))^{-1} d
\phi_{\beta\alpha} (V) \\
&= & du_\alpha (V) \phi_{\beta\alpha} (x) + u_\beta (x)
(\phi_{\beta\alpha} (x))^{-1} d \phi_{\beta\alpha} (V).
\end{eqnarray*}
Now define $\omega_\alpha (V)= \omega (d u_\alpha (V))$.  So acting
with $\omega$ on both sides we get
\begin{eqnarray*}
\omega_\beta (V) & = & \omega (du_\alpha (V) \phi_{\beta\alpha} (x)) +
\omega (u_\beta (x) (\phi_{\beta\alpha} (x))^{-1}  d \phi_{\beta\alpha}
(V) \\
& = & {\rm ad} (\phi_{\beta\alpha}^{-1} (x)) \omega_\alpha (V) +
\omega (u_\beta (x) (\phi_{\beta\alpha} (x))^{-1}  d \phi_{\beta\alpha}
(V).
\end{eqnarray*}
Now $(\phi_{\beta\alpha} (x))^{-1}  d \phi_{\beta\alpha} (V) =A \in
\mathfrak{g}$ and $u_\beta (x) A $ is the vector corresponding to the
fundamental vector field $\sigma (A)$ at $u_\beta (x)$.  Hence we have
$$\omega_\beta = {\rm ad} (\phi_{\beta\alpha}^{-1}) \omega_\alpha + 
\phi_{\beta\alpha}^{-1} d \phi_{\beta\alpha}.$$
This is to be compared with our formula from Lecture 1:
$$A'_\mu = S A_\mu  S^{-1}  - \tfrac{i}{g} \partial_\mu
S\,S^{-1}. $$
In fact, since $\phi_{\beta\alpha}$ goes from patch $\alpha$ to patch
$\beta$, and $S$ goes from unprimed to primed patch, we should
re-write this last formula by using $W=S^{-1}$:
$$ A'_\mu = W^{-1} A_\mu W + \tfrac{i}{g} W^{-1} \partial_\mu W, $$
which, apart from the  physical dimensional factor $\tfrac{i}{g} $, is
exactly the same as the first formula.

[{\bf Reference.} Tanjiro Okubo: Differential Geometry, Marcel Dekker
Inc., New York and Basel, 1987.]

In particular, we see that $\omega$ which is a 1-form {\underline{on
$P$}} can be replaced by a collection of 1-forms $\omega_\alpha$
{\underline{on $X$}}.  In the special case when the
principal bundle is trivial, we need only one such 1-form on $X$.  This is
the usual case in physics.

\medskip

A connection $\omega$ on $P$ defines a decomposition of the tangent
space at each point $u \in P$ into a {\em vertical} and a {\em
horizontal} subspace:
$$ T_u={\mathcal V}_u \oplus {\mathcal H}_u, $$
where ${\mathcal V}_u$ consists of all those tangent vectors which are
tangent to the fibre through $u$, and ${\mathcal H}_u$ those tangent
vectors annihilated by $\omega$.

\medskip

{\noindent}{\bf Definition.}\ \ {\it We can now define the {\em
exterior covariant derivative} $D\eta$ of a $p$-form $\eta$ by
$$ D\eta (V_1,\ldots,V_{p+1}) = (d \eta) (hV_1,\ldots,hV_{p+1}), $$
where $hV_i$ denotes the horizontal component of $V$.}

\medskip

{\noindent}{\bf Definition.}\ \ {\it The {\em curvature} 2-form
$\Omega$ of the connection $\omega$ is defined by $\Omega=D\omega$.}

\medskip

{\noindent}{\bf Theorem.}\ \ (The structure equation of Cartan)
$$ d\omega (X,Y) = -\half [\omega(X), \omega(Y)] + \Omega (X,Y), \ X,Y
\in T_u (P),\ u \in P. $$

\medskip

{\noindent}{\bf Theorem.}\ \ (Bianchi identity)
$$D\Omega =0. $$

\medskip

{\noindent}{\bf Definition.}\ \ {\it We say that a connection $\omega$
is {\em flat} if \ $\Omega=0$.}

\medskip

{\noindent}{\bf Definition.}\ \ {\it One can also define {\em local
curvature forms}:}
$$ \Omega_\alpha =D \omega_\alpha.$$

\medskip

It can then be shown that the following patching condition holds:
$$ \Omega_\beta = {\rm ad} (\phi^{-1}_{\beta\alpha}) \Omega_\alpha. $$

\bigskip

{\noindent}\framebox{Dictionary 2}

\smallskip

\begin{tabular}{rcl}
connection & $\longleftrightarrow$ & gauge potential\\
curvature & $\longleftrightarrow$ & gauge field
\end{tabular}

\bigskip

{\noindent}{\bf Translation.}\ \ (The structure equation of Cartan)
$$F_{\mu\nu} =\underbrace{\partial_\nu A_\mu - \partial_\mu
A_\nu}_{\mbox{\rm ``curl''}} +\ 
ig\;[A_\mu,A_\nu]. $$
(This formula appeared in Lecture 1.)

\medskip

{\noindent}{\bf Translation.}\ \ (Bianchi identity)
$$ D_\mu F_{\nu\rho} + D_\nu F_{\rho\mu} + D_\rho F_{\mu\nu} =0, $$
where $D_\mu F_{\nu\rho}= \partial_\mu F_{\nu\rho}
-ig[A_\mu,F_{\nu\rho}].$

\subsection{Bundle reductions}
\hspace*{\parindent}Let $P$ be a principal bundle with structure 
group $G$ and
let $H \subset G$ be a subgroup.  We say that $P$ is {\em reducible} 
to $H$ if there exists an open cover of $X$ such that all the
transition functions $\phi_{\beta\alpha}$ take value in $H$.  Then a
{\em reduced subbundle} or {\em reduction} $Q$ has as total space $Q
\subset P$ such that $u,v \in Q \Leftrightarrow \exists a \in H$ such
that $v=ua$.  

Given a reduction we have the associated bundle $E$ with fibre $G/H$,
structure group $G$, and total space  $E=P/H$.  Then we have

\medskip

{\noindent}{\bf Proposition.}\ \ {\it There exists a 1--1
correspondence between sections $\sigma \colon X \to E$ and reductions
$Q$.} [See Steenrod's book, in the Bibliography.]

\medskip

When we consider a connection $\omega$ on $P$, it may happen that
$\omega$ when restricted to $Q$ takes value in $\mathfrak{h}$ the Lie
algebra of $H$.   Equivalently, for all $u \in Q,\ {\mathcal H}_u (P)$
is tangent to $Q$.  In this case, we say that the connection is {\em
reducible} to $H$.  

\bigskip

{\noindent}\framebox{Dictionary 3}

\smallskip

\begin{tabular}{rcl}
bundle reduction & $\longleftrightarrow$ & symmetry breaking\\
section $\sigma \colon X \to E$ & $\longleftrightarrow$ & Higgs fields
\end{tabular}

\bigskip

{\noindent}{\bf Examples.}\ \ To be given later.

\subsection{Holonomy and loop space variables}
\hspace*{\parindent}Consider a principal bundle 
$P \stackrel{\pi}{\to} X$.  Let $\xi (s),\
s=0 \to 2\pi$, be a piecewise differentiable curve in $X$.  Then a
{\rm horizontal lift} of $\xi(s)$, denoted by $\xi^* (s)$, is  a curve
in $P$ such that $\pi (\xi^* (s)) = \xi (s)$, and all its tangent
vectors are horizontal.  Through any point $u \in P$ such that
$\pi(u)=\xi(0)$ there is a unique horizontal lift of $\xi (s)$ which
starts at $u$.

Suppose now we are given a curve $\xi(s)$ in $X$ starting from $x_0$
and ending in $x_1$.  Let $u_0$ be an arbitrary point in $\pi^{-1}
(x_0)$, and consider the horizontal lift $\xi^* (s)$ of $\xi(s)$
through $u_0$.  Let $u_1$ be its end-point, so that we have $\pi (u_1)
= x_1$.  Thus the horizontal lift defines a map $\pi^{-1} (x_0) \to 
\pi^{-1} (x_1)$ which we call the {\em parallel transport} of the
fibre above $x_0$ to the fibre above $x_1$.  It can be proved that (i)
this map is an isomorphism $\pi^{-1} (x_0) \stackrel{\sim}{\to} 
\pi^{-1} (x_1)$ and that (ii) it is independent of the parametrization
of the curve $\xi(s)$.  [Reference: Kobayashi and Nomizu, p.70.]

Suppose next that the curve $\xi(s)$ in $X$ is closed, that is
$x_1=x_0$.  Then the parallel transport is an isomorphism of $\pi^{-1}
(x_0)$ to itself.  By considering all piecewise differentiable closed
curve through $x_0$, it is easy to see that the parallel transports
form a group of automorphisms of $\pi^{-1} (x_0)$, called the {\em
holonomy group} $\Phi (u_0)$ through $u_0$, which can be identified
with a subgroup of the structure group $G$.  It can be shown that if
$X$ is connected, then all holonomy groups through any given $u \in P$
are conjugate to one another and are hence isomorphic.  Therefore, if
we are concerned only with the abstract holonomy group for a given
connection and not which particular subgroup of $G$ it is, then we can
omit the reference to $u_0$.  We can thus consider simply the {\em 
holonomy} $\Phi (C)$ of a closed curve $C$ in $X$ as an element of $G$.

\bigskip

{\noindent}\framebox{Dictionary 4}

\smallskip

holonomy  $\longleftrightarrow$  phase factor\\
(Compare Lecture 1.)

\medskip

{\noindent}{\bf Note.}\ \ The curvature at a given point can again be
thought of as the holonomy of an infinitesimal closed looop through
that point.

\subsubsection*{Flat connections}
\hspace*{\parindent}Suppose $X$ is connected.  If $X$ is not 
simply connected, then a flat
connection may give rise to nontrivial holonomy.  In fact, there is a
1--1 correspondence:
$$\left\{ \begin{array}{c}
{\rm gauge\ equivalence\ classes}\\ {\rm of\ flat\ connections}
\end{array} \right\} \stackrel{1-1}{\longleftrightarrow}
\left\{ \begin{array}{c}
{\rm conjugacy\ classes\ of\ irreducible}\\ {\rm representations\ of}\
\pi_1(X) \to G   \end{array} \right\}  $$

\bigskip

Let $\Omega^1 X$ be the space of closed piecewise differentiable loops
in $X$, called the {\em loop space} of $X$.  For convenience, we
always consider loops starting and ending at a fixed point $x_0$.
Given any connection $\omega$, the holonomy $\Phi$ defines a map
$$ \Phi \colon \Omega^1 X \to G  $$
which satisfies the {\em composition law}.  This means that given two
loops $C_1$ and $C_2$, we can compose them to give a third loop $C_1
\circ C_2$, by first going round $C_1$ for $s=0 \to \pi$ and then
going round $C_2$ for $s=\pi \to 2\pi$, for instance.  Then we have
$$ \Phi (C_1 \circ C_2) = \Phi (C_2)\,\Phi (C_1), $$
where the product on the right-hand side is group multiplication.

The converse problem: given a map $\Phi \colon \Omega^1 \to
G$ satisfying the composition law (and some other obvious conditions),
can one define a connection of which $\Phi$ is its holonomy?  This
problem has great importance for Yang--Mills theory and is to a large
extent solved.  However, the known proofs are all hard and not totally
rigorous.  We shall not present them here.

It is, however, interesting to see why the problem is of importance to
Yang--Mills theory.

I mentioned in Lecture 1 that Yang proved that the Dirac phase factor
describes gauge theory exactly, in the sense that there is a 1--1
correspondence between
$$\{ \Phi(C) \} \longleftrightarrow \{ {\rm physical\ configurations\
YM} \}.  $$
This is in contrast to the variables $A_\mu$ which depend on gauge
(which means coordinate bundle in the language of this Lecture), or
the variables \fdown, which cannot distinguish {\em all} physically
different situations.  However, we have to be extremely careful in not
confusing this with the concept of {\em redundancy}.  A moment's
thought will tell us that not all maps
$$\Phi \colon \Omega^1 \to G $$
come from a connection, and further study will reveal the fact that
adding the composition law (and certain other obvious conditions) is
not enough.  In other words, we have to impose {\em constraints} on
the variables $\Phi$.

The situation can be summarized as follows.  If we use the variables
$A_\mu$, then the physically relevant objects are equivalence classes
of $A_\mu$ under gauge equivalence.  If we use the variables
$\Phi(C)$, then we have to find the relevant subset by imposing
constraints.  Depending on the problem at hand, it is sometimes easier
to deal with the quotient space (the case of $A_\mu$) or to deal with a
subspace (the case of $\Phi(C)$).

I shall now give a rough non-rigorous description of what the main
constraint is.  Given a closed loop $C$, we can make a
$\delta$-function variation at any point $s$ on the curve.  As the
height of this $\delta$-function goes to zero, we can then define the
{\em loop derivative} $\delta_\mu (s)$ of any loop-dependent
quantity.  Remembering that $\Phi(C)$ is an element of $G$, we define
the Lie algebra-valued quantity $F_\mu (C,s)$ as the logarithmic
derivative of $\Phi(C)$:
$$ F_\mu (C,s) = \frac{i}{g}\;\Phi^{-1} (C)\;(\delta_\mu (s) \Phi
(C)). $$

\begin{figure}[ht]
\centering
\input{loopderiv.pstex_t}
\caption{Sketch of loop derivative.}
\label{loopderiv}
\end{figure}
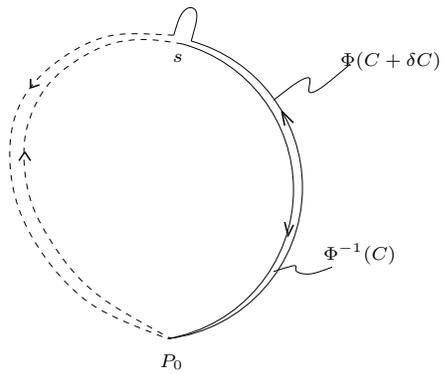

This can be illustrated by the sketch in Figure \ref{loopderiv}.  In 
asmuch as this quantity carries the phase factor from one loop to 
a neighbouring loop, it is like an infinitesimal phase transport and
can indeed be regarded as some sort of ``connection'' in a coordinate
bundle over $\Omega^1 (X)$.  We can even go further and consider {\em
its holonomy}, this time of a closed loop in $\Omega^1 (X)$, which
means a closed surface $\Sigma$ in $X$.

The constraint we are after is that this connection $F_\mu(C,s)$ be
{\em flat}.  We note that this does not necessarily imply that the
corresponding holonomy is trivial.   Any such nontrivial holonomy can
be interpreted as a nonabelian {\em magnetic monopole} charge, as we
shall see.

\clearpage

\section{Gauge group and charges}

\subsection{Locally isomorphic semi-simple Lie groups}
\hspace*{\parindent}So far, we have been rather vague about the 
exact gauge group that
occurs in a particular Yang--Mills theory.  A particularly
interesting, but often neglected, aspect is the different choices of
Lie groups which correspond to the same Lie algebra.  In asmuch as the
Lie algebra can be identified (as a vector space) with the tangent
space at the identity, it is clear that groups which are locally
isomorphic (that is, quotients by discrete subgroups) have the same
Lie algebra.  In fact, for semi-simple groups, among all locally
isomorphic groups there is one which is simply connected and which is
the universal cover of all the others.  These latter can then be
obtained from the universal cover group by factoring out by various
subgroups of its (discrete) centre.

Some examples will make this clear.

\begin{enumerate}
\item Consider the matrix groups $SO(3)$ and $SU(2)$.  The Lie algebra
of $SO(3)$ consists of $3 \times 3$ skew matrices, and we can choose as
generators:
$$Y_1=\left( \begin{array}{rrr}
0 & 0 & 0\\ 0 & 0 & 1 \\ 0 & -1 & 0 \end{array} \right), \quad 
Y_2=\left( \begin{array}{rrr}
0 & 0 & 1\\ 0 & 0 & 0 \\ -1 & 0 & 0 \end{array} \right), \quad 
Y_3=\left( \begin{array}{rrr}
0 & 1 & 0\\ -1 & 0 & 0 \\ 0 & 0 & 0 \end{array} \right), $$
which satisfy the commutation relations
$$ [Y_i,Y_j] = \epsilon_{ijk} Y_k. $$
The Lie algebra of $SU(2)$ consists of the trace-free skew hermitian
$2 \times 2$ matrices, and we can choose as generators:
$$X_i = - \half i \sigma_i, $$
where $\sigma_i$ are the Pauli matrices
$$ \sigma_1 = \left( \begin{array}{rr}
0 &1\\ 1 & 0 \end{array} \right), \quad
\sigma_2 = \left( \begin{array}{rr}
0 &-i\\ i & 0 \end{array} \right), \quad
\sigma_3 = \left( \begin{array}{rr}
1 &0\\ 0 & -1 \end{array} \right). $$
They satisfy
$$ [X_i,X_j] = \epsilon_{ijk} X_k. $$
Hence the two Lie algebras are isomorphic.  The groups are not
isomorphic, only locally isomorphic.  In fact, there is a 2--1 map
$$ SU(2) \longrightarrow SO(3)$$
in such a way that $SU(2)$ is a double cover of $SO(3)$.  This map can
easily be worked out explicitly ({\em Exercise}).

More generally, we can think of $SU(2)$ as the unit sphere $S^3$ in
$\bbr^4$, by identifying $SU(2)$ with unit quaternions ($SU(2) \cong
Sp(1)$).  Then the 2--1 map corresponds to identifying antipodal
points of $S^3$.

Furthermore, this discrete quotient is by the centre $\bbz_2$ of
$SU(2)$, and this discrete group is the fundamental group of $SO(3)$.
\item Very similar considerations apply to $SU(N)$, with centre
$\bbz_N$.  In the case of $SU(3)$, we have altogether 2 locally
isomorphic groups: $SU(3)$ and $SU(3)/\bbz_3$.  In the case of $SU(6)$
we have 4: $$SU(6), SU(6)/\bbz_2, SU(6)/\bbz_3, SU(6)/\bbz_6,$$ where
$\bbz_2$ and $\bbz_3$ are subgroups of the centre $\bbz_6$.
\item The group $SU(2) \times U(1)$ is a double cover of $U(2)$, the
covering map being given by multiplication as follows.  First embed
$U(1) \to U(2)$ by
$$ e^{i\alpha} \mapsto \left( \begin{array}{cc}
e^{i\alpha} & 0\\ 0 & e^{i\alpha} \end{array} \right).  $$
Then
\begin{eqnarray*}
SU(2) \times U(1) & \stackrel{2-1}{\longrightarrow} & U(2) \\
\left( \left( \begin{array}{cc}
a & b \\ c & d \end{array} \right), \left( \begin{array}{cc}
e^{i\alpha} & 0\\ 0 & e^{i\alpha} \end{array} \right) \right) &
\longmapsto & \left( \begin{array}{cc}
a e^{i\alpha} & b e^{i\alpha} \\ c e^{i\alpha} & d e^{i\alpha}
\end{array} \right).\\
\| \qquad \qquad \| \qquad \qquad&&\\
f \qquad \qquad g \qquad \qquad&&  
\end{eqnarray*}
We see immediately that $(f,g)$ and $(-f,-g)$ have the same image in
$U(2)$. 
\end{enumerate}

\subsection{Specification of the gauge group}
\hspace*{\parindent}Recall (Lecture 1) that gauge invariance comes 
about as the invariance
of the wavefunction of a charged particle under the action of a group
$G$, so that to specify $G$ one has to examine {\em all} the charged
particles occurring in the theory, in other words, its {\em spectrum}.

Start with electromagnetism.  Under a phase rotation, $\psi \mapsto
e^{ie \Lambda} \psi$, so that we can parametrize the circle group
$U(1)$ corresponding to the phase by $[0, 2\pi/e]$.

Next suppose there are charges $e_1,\ldots,e_k$ in the theory; then 
$\psi_r \mapsto e^{ie_r \Lambda} \psi_r,\ r=1,\ldots,k$.  If the
charges are commensurate, that is, if there exist $e$ 
and integers $n_r$ such that
$$ e_r=n_re,\quad r=1,\ldots,k,$$
then again we can parametrize the $U(1)$ by $[0,2\pi/e]$, because if
$\Lambda$ changes by any integral multiple of $2\pi/e$, the
wavefunctions corresponding to {\em all} the charges will be
unchanged.  If, however, there is at least one pair of charges whose
ratio is irrational, then we no longer have $U(1)$ as a gauge group.
In fact, {\em charge quantization} is equivalent to having $U(1)$ as
the gauge group of electromagnetism.

On the other hand, if we consider pure electromagnetism without
charges, then the only relevant gauge transformation are those of
$A_\mu$:
$$ A_\mu \mapsto A_\mu + \partial_\mu \Lambda, $$
so that the group will just be the real line given by the scalar
function $\Lambda (x)$.

Similar considerations apply to nonabelian theory.  In the vast
majority of cases, from the physics point of view, one knows the Lie
algebra, and then one needs to inspect the spectrum to get the correct
Lie group.  One must bear in mind that this implies, for any given
Yang--Mills theory, that if in future the spectrum is changed for any
reason, one may have to consider another Lie group instead.

Consider first a {\em pure} Yang--Mills theory without charges, so
that the only gauge transformation one needs to consider is on the
gauge potential $A_\mu (x)$:
$$A_\mu \mapsto S\,A_\mu\,S^{-1} - \frac{i}{g} \partial_\mu
S\,S^{-1}. $$
Let $G$ be the universal cover of all the groups corresponding to the
given Lie algebra.  Then the effects on $A_\mu$ of $S$ and $\gamma S$,
where $\gamma$ is an element of the centre $Z$ of $G$, are identical.
Hence the correct gauge group must be $G/Z$, which is in an obvious
sense the smallest of all the possible groups.  So in the example (1)
we considered, the group is $SO(3)$ and not $SU(2)$.

On the other hand, if the $\mathfrak{su} (2)$ theory contains
particles with a 2-component wave function $\psi=\{\psi_i,\ i=1,2\}$,
then
$$\psi \mapsto S \psi,\quad S \in SU(2)$$
and the effect of $S$ and $-S$ are not identical.  Hence in this case
the correct gauge group is indeed $SU(2)$ and not $SO(3)$.

These considerations can also be cast in terms of representations.
Charged particles in a Yang--Mills theory are in certain
representations of the gauge group.  What we are saying is the known
result that the collection of all representations determines the
group.  In the above case, the gauge potential is in the 3-dimensional
adjoint representation and the 2-component $\psi$ is in the
2-dimensional spinor representation.   In the absence of the spinor
representation, the group is $SO(3)$, but when spinors are present, the
group must be $SU(2)$.  This representation theory is entirely
identical to the theory of spin and angular momentum in quantum
mechanics.
  
We now apply these considerations to the Yang--Mills theories
occurring in particle physics.  For these we suppress the gauge
couplings $g$ for convenience.

\begin{enumerate}
\item {\underline{Strong interaction}}.\ \ Because we postulate the
existence of quarks, which are in the 3-dimensional fundamental
representation of $SU(3)$, we conclude the gauge group is indeed
$SU(3)$.
\item {\underline{Electroweak interaction}}.\ \ The particles are of 2
types (where `flavour' means `weak isospin'):
\begin{enumerate}
\item $SU(2)$ flavour doublets with half-integral weak
hypercharge---$T_3=\pm \half,\ Y=\tfrac{k}{2},\ k$ odd;
\item $SU(2)$ flavour singlets or triplets with integral weak
hypercharge---$T_3 = 0,1,\ Y=k$.
\end{enumerate}
Under a gauge transformation generated by the generators $T_3$ and
$Y$, we have for the two types of particles:
$$\begin{array}{rcccl}
(a)\quad (\exp 2\pi i T_3)\,\psi & = & (\exp i\pi)\,\psi & = & - \psi \\
(\exp 2\pi i Y)\,\psi & = & (\exp i\pi)\,\psi & = & - \psi \\
(b)\quad (\exp 2\pi i T_3)\,\psi & = & (\exp i\pi)\,\psi & = & \psi \\
(\exp 2\pi i Y)\,\psi & = & (\exp i\pi)\,\psi & = & \psi,
\end{array} $$
so that the resultant action of $T_3+Y$ in {\em both} cases is the
identity.  Hence we conclude that in the group $SU(2) \times U(1)$ we
should identify pairs $(f,g) \equiv (-f,-g)$, so that the correct
gauge group for electroweak theory is $U(2)$.

However, if in future we either {\em discover} or {\em postulate} more
particles, e.g.
\begin{enumerate}
\item[(c)] $SU(2)$ flavour doublets with integral weak hypercharge,
and/or
\item[(d)] $SU(2)$ flavour singlets or triplets with half-integral
weak hypercharge,
\end{enumerate}
{\em then} the effect of $(f,g)$ and $(-f,-g)$ on these particles are
distinct, and in that case the correct group is $SU(2) \times U(1)$.
\item {\underline{Standard model}}.\ \ Similar considerations of the
known/postulated spectrum, as given in Lecture 2, will show that we
should have a 6-fold identification in $SU(3) \times SU(2) \times
U(1)$, where the following 6 triplets should be identified:
$$(c,\!f,\!y), \!(cc_1,\!f,\!yy_1), \!(cc_2,\!f,\!yy_2), 
\!(c,\!f\tilde{f},\!y\hat{y}), 
   \!(cc_1,\!f\tilde{f},\!y\hat{y}y_1), \!(cc_2,\!f\tilde{f},
\!y\hat{y}y_2), $$
where $c, f$ and $y$ are elements respectively of $SU(3)$, $SU(2)$
and $U(1)$, with:
\begin{eqnarray*}
c_r & = & \exp \frac{2 \pi i r}{\sqrt{3}} \lambda_8, r = 1, 2;\\
\tilde{f} & = & \exp 2 \pi i T_3; \\
y_r & = & \exp 4 \pi i r Y, r = 1, 2; \\
\hat{y} & = & \exp 6 \pi i Y,
\end{eqnarray*}
and
$$ \lambda_8 = \frac{1}{\sqrt{3}} \left( \begin{array}{rrr}
1 & 0 & 0 \\ 0 & 1 & 0 \\ 0 & 0 & -2 \end{array} \right), \quad T_3=
\left( \begin{array}{rr}
1 & 0 \\ 0 & -1 \end{array} \right), \quad Y=\frac{1}{6},  $$
with obvious embeddings in $SU(5)$ and abuse of notation (same symbol
for generators of different representations).

\end{enumerate}

\subsection{Charges and monopoles}
We have known for a long time what the electric charge is.  There
are several equivalent ways of defining or describing it.  For our
purpose here, we shall consider it as giving a nonvanishing right
hand side to Maxwell's equation: $$ \partial_\nu F^{\mu\nu} =
-j^\mu, $$ where the current $j^\mu$ is given in one of two
ways\footnote{In the above I have introduced the gamma matrices
$\gamma^\mu$, which are important ingredients in Dirac's theory of the
spin $\half$ particles and which provide a prosaic way of using
Clifford algebras.  For lack of time I shall not expand into the
subject, but they are treated in depth in Hijazi's lectures.  See also
the lectures of Langmann.}:
$$j^\mu = \left\{\begin{array}{ll} e \int d\tau (dY^\mu/d \tau) & {\rm
classical} \\ e \bar{\psi} \gamma^\mu \psi & {\rm quantum}
\end{array} \right.$$ 
The quantity $e$ here in fact plays a double role:
\begin{enumerate}
\item[(a)] it is the electric charge---it determines how the charge
interacts with the field, and
\item[(b)] it is the coupling constant---it fixes the strength of
the interaction. \end{enumerate}

In Yang--Mills theory, a non-abelian electric charge (sometimes
referred to generically as a ``colour electric charge'') can also
be thought of as giving a non-vanishing right hand side to the
Yang--Mills equation:  $$ D_\nu F^{\mu\nu} = -j^\mu, $$ where
$j^\mu= g \bar{\psi} \gamma^\mu \psi$ for the quantum particle.
[In Yang--Mills theory, one does not usually concern oneself with
classical charges.]

But here the two roles (a) and (b) are quite distinct.  The
wavefunction $\psi$ is in a certain representation of the gauge
group $G$, and {\em how} the particle interacts with the gauge
field is determined by the representation, the interaction being
given by the covariant derivative.  The coupling constant $g$, on
the other hand, is the numerical factor which fixes the strength
of the interaction.

We see that both abelian and nonabelian electric charges occur as
nonvanishing {\em currents}.  In order to make a practical
distinction between these and the topological charges we shall
discuss next, let us call them {\em electric charges}, or simply
{\em charges} when there is no confusion.

There is another type of charges called {\em monopoles}.  They are 
typified by the
magnetic monopole as first discussed by Dirac in 1931.

To understand them better, let us look at Maxwell's equations both
in the 3-vector and 4-vector notations:
$$\begin{array}{ll} \left.
\begin{array}{r} {\rm div}\ \mathbf{E} = \rho \\ {\rm curl}\ \mathbf{B} -
\partial \mathbf{E}/\partial t = \mathbf{J} \end{array} \right\}
 & \partial_\nu F^{\mu\nu} =-j^\mu \\ 
& \\
\left.
\begin{array}{r} {\rm div}\ 
\mathbf{B} =0 \\ {\rm curl}\ \mathbf{E} + \partial \mathbf{B}/\partial
t =0 \end{array} \right\} & \partial_\nu \fstarup =0.
\end{array} $$
Here the charge density $\rho$ and the electric
current $\mathbf{J}$ together form the 4-current $j^\mu$.  We also
define the {\em dual} field tensor \fstarup\ by
$$\fstarup = -\half \epsilon^{\mu\nu\rho\sigma}
F_{\rho\sigma},$$ 
where $\epsilon^{\mu\nu\rho\sigma}$ is the totally skew symbol 
with the convention that $\epsilon^{0123}=-1$.

At the {\em classical} particle level, these equations simply tell
us the experimental fact that magnetic charges, called {\em magnetic
monopoles}, do not exist in nature.  If, on the other hand, we are
concerned with {\em quantum} particles, then the Bohm--Aharonov
experiment (Lecture 1) tells us that we have to introduce the {\em
vector potential} $A_\mu$ bearing the relation with the field
\fdown\ as $$ \fdown = \partial_\nu A_\mu - \partial_\mu A_\nu. $$
Simple algebra will tell us that this implies $\partial_\nu \fstarup
=0$ as above.  Hence we conclude that: 
$$ \exists\ {\rm monopole}
\Longrightarrow A_\mu\ {\rm cannot\ be\ well\ defined\
everywhere.}$$ 
The result is actually stronger.  Suppose there
exists a magnetic monopole at a certain point in spacetime, and
without loss of generality we shall consider a static monopole. If
we surround this point by a (spatial) 2-sphere $\Sigma$, then the
magnetic flux out of the sphere is given by 
$$ \int\!\!\!\int_\Sigma
\mathbf{B} \cdot \mbox{\boldmath$d \sigma$} =
\int\!\!\!\int_{\Sigma^N} 
\mathbf{B} \cdot \mbox{\boldmath$d \sigma$}
 + \int\!\!\!\int_{\Sigma^S} \mathbf{B} \cdot \mbox{\boldmath$d 
\sigma$}, $$ 
where $\Sigma^N$ and $\Sigma^S$ are the northern and southern
hemispheres intersecting at the equator $S$. By Stokes' theorem
since \fdown\ has no components $F_{0i}=E_i$, 
\begin{eqnarray*}
\int\!\!\!\int_{\Sigma^N} \mathbf{B} \cdot \mbox{\boldmath$d \sigma$} 
& = & \oint_S
\mathbf{A} \cdot \mathbf{ds} \\
\int\!\!\!\int_{\Sigma^S} \mathbf{B}
\cdot  \mbox{\boldmath$d \sigma$}& = & \oint_{-S} \mathbf{A} \cdot 
\mathbf{ds},
\end{eqnarray*} 
where $-S$ means the equator with the opposite
orientation.  Hence $ \oint_S + \oint_{-S} =0$.  But this
contradicts the assumption that  there exists a magnetic monopole
at the centre of the sphere.  Hence we see that if a monopole
exists, then $A_\mu$ will have {\em at least} a string of
singularities leading out of it. This is the famous {\em Dirac
string}.

The more mathematically elegant way to describe this is
that the principal bundle corresponding to electromagnetism with a
magnetic monopole is {\em nontrivial}, so that the gauge potential
$A_\mu$ has to be {\em patched} (i.e.\ related by transition
functions). [Recall the collection of local 1-forms
$\omega_\alpha$.] Consider the example of a {\em static monopole}
of magnetic charge $\tilde{e}$.  For any (spatial) sphere $S_r$ of
radius $r$ surrounding the monopole, we cover it with two patches
$N,S$: 
$$\begin{array}{lll} (N)\ \colon & 0 \leq \theta < \pi, & 0
\leq \phi \leq 2\pi \\ (S)\ \colon & 0 < \theta \leq \pi & 0 \leq
\phi \leq 2\pi, \end{array} $$ 
and define in each patch:
$$\begin{array}{lll} A_1^{(N)} = \frac{\tilde{e} y}{r(r+z)}, &
A_2^{(N)} = -\frac{\tilde{e} x}{r(r+z)}, & A_3^{(N)} =0; \\
A_1^{(S)} = -\frac{\tilde{e} y}{r(r-z)}, & A_2^{(S)} =
\frac{\tilde{e} x}{r(r-z)}, & A_3^{(S)} =0. \end{array} $$  
In the
overlap (containing the equator), $A^{(N)}$ and $A^{(S)}$ are
related by a gauge transformation 
$$A_i^{(N)} - A_i^{(S)} =
\partial_i \Lambda, \quad \Lambda= 2 \tilde{e} {\tan}^{-1} (y/x)= 
2 \tilde{e} \phi. $$  
Notice
that $A_i^{(N)}$ has a line of singularity along the negative
$z$-axis (which is the Dirac string in this case).  Similarly for
$A_i^{(S)}$.

Furthermore, the corresponding field strength is: 
$$ \mathbf{E}
=0, \quad \mathbf{B} = \tilde{e} \mathbf{r} /r^3. $$  
If we now
evaluate the `magnetic flux' out of $S_r$, we have 
$$
\int\!\!\!\int_{S_r} \mathbf{B} \cdot  \mbox{\boldmath$d \sigma$}
= \oint_{\rm equator}
(A_\mu^{(N)} - A_\mu^{(S)}) dx^\mu = 4\pi \tilde{e}, $$ 
in other words, in the presence of a magnetic monopole the last two
Maxwell's equations are modified:  
$$ \left. \begin{array}{r}
{\rm div}\ \mathbf{B} = \tilde{\rho} \\ {\rm curl}\ \mathbf{E} + \partial
\mathbf{B}/\partial t = \tilde{\mathbf{J}} \end{array} \right\} \quad
\partial_\nu {}^*\!F^{\mu\nu} = - \tilde{\jmath}^\mu,$$ 
with
$$\tilde{\jmath}^\mu = \left\{ \begin{array}{ll} \tilde{e} \int d
\tau \frac{d Y^\mu}{d \tau} \delta (x-Y(\tau)) & {\rm classical}
\\ \tilde{e} \bar {\psi} \gamma^\mu \psi & {\rm quantum.}
\end{array}\right. $$

How are the charges $e$ and $\tilde e$ related?  Well, the gauge
transformation $S=e^{ie\Lambda}$ relating $A_\mu^{(N)}$ and
$A_\mu^{(S)}$ must be well-defined, that is, if one goes round
the equator once: $\phi=0 \to 2\pi$, one should get the {\em same}
$S$.  This gives:  
$$ 2e \tilde{e} (2\pi) = 2n\pi, \quad n \in
\bbz\quad \Longrightarrow $$ 
$$  e \tilde{e} = n/2. \quad ({\rm Dirac\
quantization\ condition}) $$  In particular, the unit electric and
magnetic charges are related by $$e \tilde{e} = 1/2. $$  So in
principle, just as in the electric case, where we could have
charges $e,2e,\ldots$, here we could also have magnetic charges of
$\tilde{e}, 2\tilde{e},\ldots$.  In other words, both charges are
{\em quantized}.

Another way to look at this fact is to consider the classification
of principal bundles over $S^2$.  The reason for these topological
2-spheres is that we are interested in {\em enclosing} a point
charge.  For a nontrivial bundle, the patching is given by a
function $S$ defined in the overlap (the equator), in other words,
a map $S^1 \to U(1)$.  What this amounts to is a closed curve in
the circle group $U(1)$.  Now curves which can be continuously
deformed into one another cannot give distinct fibre bundles, so
that one sees easily that there exists a 1--1 correspondence
between 
\begin{center} 
\{principal $U(1)$ bundles over
$S^2$\} \quad $\stackrel{1-1}{\longleftrightarrow}$ \\ 
\{homotopy classes of closed curves
in $U(1)$\}$=\pi_1 (U(1)) \cong \bbz.$
\end{center}
Hence we
recover Dirac's quantization condition.

So for electromagnetism, there are two equivalent ways of defining
the magnetic charge: \begin{enumerate} \item $\partial_\nu
{}^*\!F^{\mu\nu} = - \tilde{\jmath}^\mu \varpropto n \tilde{e} \ne
0$
\item an element of $\pi_1 (U(1)) \cong \bbz$.  \end{enumerate}
We also note that both give us the fact that these charges are (A)
discrete (quantized) and (B) conserved (invariant under continuous
deformations).

We now want to define the magnetic monopoles in the nonabelian
case.  For simplicity, these are sometimes referred to as ``colour
magnetic monopoles''.  [Note that there is another kind of
monopole which is a solitonic solution and not a fundamental
charge as these are, as we shall explain briefly later.]

For several (subtle) reasons the obvious expression (see Table
\ref{charges}) $$D_\nu {}^*\!F^{\mu\nu} = -
\tilde{\jmath}^\mu$$ does not work.
\begin{table}
\centering
\begin{tabular}{|l|c|c|}\hline  
& Charges & Monopoles \\ \hline
 abelian & $\partial_\nu F^{\mu\nu} = - j^\mu$ & $ \partial_\nu 
{}^*\!F^{\mu\nu} = - \tilde{\jmath}^\mu $ \\ 
nonabelian &
$D_\nu F^{\mu\nu} = - j^\mu$ & ? \\ \hline 
\end{tabular}
\caption{Definitions of charges} 
\label{charges} 
\end{table}

The quickest way to say it is that $ {}^*\!F^{\mu\nu}$
in general has no corresponding potential $\tilde{A}_\mu$ and so
cannot describe the quantum monopole.  We shall come back to this
later.

But we just saw that in the abelian case there is another
equivalent definition, and that is, a magnetic monopole is given
by the gauge configuration corresponding to a nontrivial $U(1)$
bundle over $S^2$.  This can be generalized to the nonabelian case
without any problem.  Moreover, this definition automatically
guarantees that a nonabelian monopole charge is (A) quantized and
(B) conserved.  We invoke the following classification result.

\smallskip

{\noindent}{\bf Proposition.}\ \ {\em The equivalent classes of
nontrivial $G$ bundles over $S^2$ are in 1--1 correspondence with
the elements of $\pi_1 (G)$.} [The proof is very similar to the
abelian case.  See Steenrod's book, in the Bibliography.]

\smallskip

{\noindent}{\bf Definition.}\ \ {\em A nonabelian monopole for
gauge group $G$ is given by an element of $\pi_1 (G)$.}

\subsection{Examples.}
\begin{enumerate}
\item $\pi_1 (U(1)) = \bbz$.
\item $\pi_1 (SU(N)) = 0 \Longrightarrow$ no monopoles.
\item $\pi_1 (SO(3)) = \bbz_2 \Longrightarrow \exists $ the vacuum
and one type of monopole only.  Charges can be denoted by a sign
$+$ or $-$.
\item $\pi_1 (SU(3)/\bbz_3) = \bbz_3 \Longrightarrow $ charges are
given by the cube roots of unity $1, \omega, \omega^2$.
\end{enumerate}

\vspace*{5mm}
\noindent{\bf Example of $SO(3)$ monopole charge $-$}

\medskip

\begin{figure}
\centering
\includegraphics{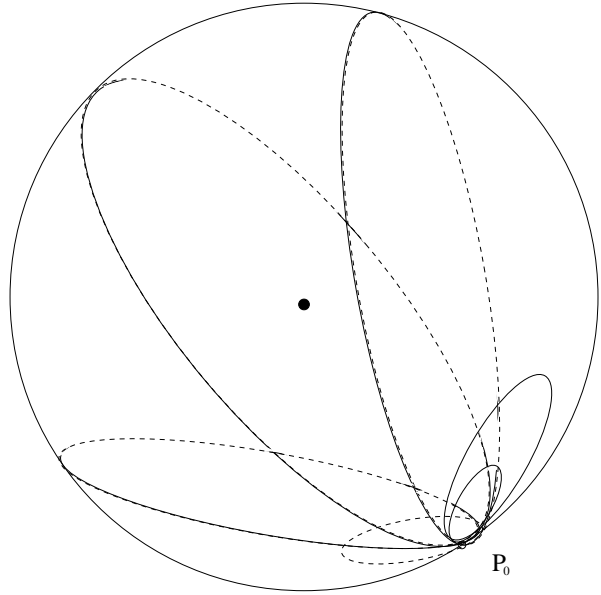}
\caption{Surface swept out by the one-parameter family of loops $\xi_t$.}
\label{looptheloop}
\end{figure}

On a 2-surface $\Sigma \simeq S^2$ enclosing the monopole, choose a
family of closed curves spanning it, as illustrated in Figure
\ref{looptheloop}. 
$$\xi_t (s) \colon s=0 \to 2\pi, \quad t=0 \to 2\pi, $$
with
$$ \xi_t (0) =\xi_t (2\pi) = P_0, \quad \xi_0 (s) = \xi_{2\pi} (s) =
P_0. $$

We shall work in $SU(2)$, so that 
\begin{center}
monopole charge $\simeq$ class of curves going half-way round
\end{center}
In other words, we consider the holonomy to be an element of $SU(2)$.

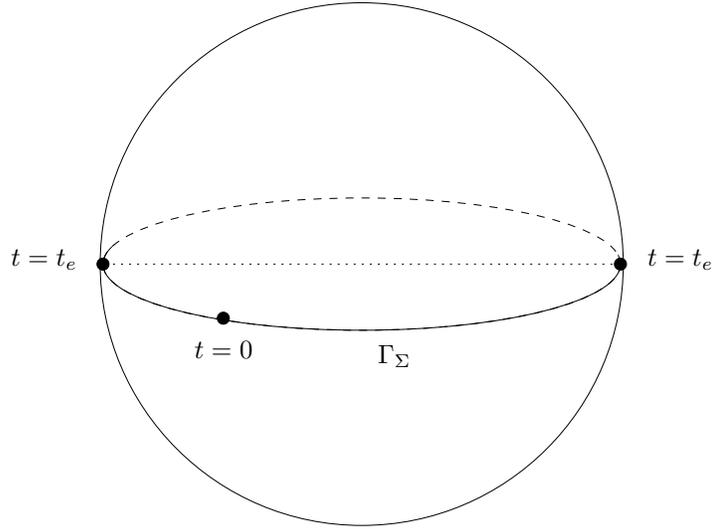
\begin{figure}
\centering
\input{holoncurv.pstex_t}
\caption{A curve representing an $SO(3)$ monopole}
\label{holoncurv}
\end{figure}

Without loss of generality, choose the base point $P_0$ to be on the
equator, corresponding to the loop $\xi_{t_e} (s)$, as in Figure
\ref{holoncurv}. 
Starting at $t=0, \Phi^N (\xi_0) =1$, the phase factor $\Phi^N
(\xi_t)$ traces out a continuous curve until it reaches $t=t_e$.  Then
one makes a patching transformation and goes over to $\Phi^S
(\xi_t)$.  From $t_e$ onwards this again traces out a continuous curve
ending in 1 for $t=2\pi$.  In order that the curve $\Gamma$ so traced
out winds round only half way in $SU(2)$ while being a  closed curve
in $SO(3)$ we must have
$$ \Phi^N (\xi_{t_e}) = - \Phi^S (\xi_{t_e}).  $$
So the holonomy of the closed loop $\xi_t$ in $\Omega \Sigma$
corresponding to the flat connection $F_\mu(\xi,s)$ is $-1$, which is
the monopole charge.

\clearpage

\section{Action principle and symmetry breaking}
\hspace*{\parindent}So far we have not discussed dynamics at all, 
that is, nothing much is
said about the time evolution of the gauge system.  These are given,
in the classical and first quantized cases, by equations of motion,
wich are normally obtained by the first variation of the functionals
of the fields called {\em actions}.  Various actions will describe
various physical systems.  We shall study some of them.

\subsection{Maxwell equations and minimal coupling}
\hspace*{\parindent}The Maxwell action is usually given as:
$$ {\mathcal A}_F^0 = - \quart \int \fdown \fup d^4 x. $$
This is the free field action, that is, it corresponds to a freely
propagating electromagnetic field.

Recall that $\fdown=\partial_\nu A_\mu - \partial_\mu A_\nu$.  Then
$$ {\mathcal A}_F^0 = - \quart \int (\partial_\nu A_\mu - \partial_\mu
A_\nu) \fup d^4 x = - \half \int (\partial_\nu A_\mu) \fup d^4 x. $$
Varying with respect to $A_\mu$ we get
$$ \partial_\nu \fup = 0. $$
This, we recall, is the first pair of Maxwell's vacuum equations in
covariant notation.  The second pair, in this situation, can be
considered as {\em kinematics}, because by the definition of \fdown\
it is an identity:
$$\partial_\rho \fdown + \partial_\mu F_{\nu\rho} + \partial_\nu
F_{\rho\mu} =0, $$
or equivalently
$$ \partial_\nu \fstarup =0, $$
with $\fstarup = - \half \epsup F_{\rho\sigma}$.

On the other hand, if we have a free (classical) particle in free
space, then we define its free action
$$ {\mathcal A}_M^0 = -m \int_Y d \tau, $$
where $\tau$ is the proper time, the integral is along the worldline
of the particle, with $\tfrac{dY^\mu}{d\tau}
\tfrac{dY_\mu}{d\tau}=-1$.  Varying with respect to the worldline
$Y^\mu (\tau)$, we get
$$ m \frac{d^2 Y^\mu}{d \tau^2} = 0.  $$

If we simply add the two actions together, we will not of course get
any interaction.  One way to introduce interaction is to add an
``interaction term'', using the {\em minimal coupling assumption}:
$$ {\mathcal A}_I = -e \int A_\mu (Y(\tau)) \frac{dY^\mu
(\tau)}{d\tau} d\tau.  $$
So we have the total action:
$$ {\mathcal A}_{\rm m.c.} = {\mathcal A}_F^0 + {\mathcal A}_M^0 +
{\mathcal A}_I.  $$
Varying with respect to $A_\mu (x)$ and $Y^\mu (\tau)$ we get
\begin{eqnarray*}
 \partial_\nu \fup (x) & = & -e \int d\tau \frac{dY^\mu
(\tau)}{d\tau} \delta^4 (x-Y(\tau)) \\
m \frac{d^2 Y^\mu}{d \tau^2} & = & -e \fup (Y(\tau)) \frac{dY_\nu
(\tau)}{d\tau}.
\end{eqnarray*}
The first is Maxwell's equation in the presence of an electric
current-density $j^\mu$, and the second is the Lorentz equation for a
charge moving in an electromagnetic field.

For the quantum particle, which we assume to be a Dirac particle
(i.e.\ with
spin $\half$), then we replace the particle action by
$$ {\mathcal A}_M^0 = \int d^4 x \bar{\psi} (x) (i \partial_\mu
\gamma^\mu -m) \psi (x), $$
and the interaction term by
$$ {\mathcal A}_I = -ie \int \bar{\psi} \gamma^\mu A_\mu \psi, $$
which on variation with respect to $A_\mu$ and $\bar{\psi}$ give the
following equations of motion:
\begin{eqnarray*}
\partial_\nu \fup (x) &=& -e \bar{\psi} (x) \gamma^\mu \psi (x) \\
(i\partial_\mu \gamma^\mu -m) \psi (x) & = & -e A_\mu (x) \gamma^\mu
\psi (x). 
\end{eqnarray*}
These are then the quantum equations, the second being the well-known 
Dirac equation.

\subsection{Yang--Mills equation}
\hspace*{\parindent}We can do the same for Yang--Mills theory.

In the free theory, we have the same action:
$$ {\mathcal A}_F^0 = - \quart \int \fdown \fup d^4 x, $$
where now $\fdown=\partial_\nu A_\mu - \partial_\mu A_\nu + ig [A_\mu,
A_\nu].$   As before, varying with respect to $A_\mu$ we get the
equations of motion, this time the
Yang--Mills equation:
$$D_\nu \fup = 0. $$ 
Here the covariant derivative is 
$D_\mu = \partial_\mu -ig[A_\mu, \ ]$.

Again, in the presence of a colour electrically charged Dirac particle
(a `quark' for example), one can introduce an interaction term by
the hypothesis of minimal coupling:
$$ {\mathcal A}_I = -ig \int \bar{\psi} \gamma^\mu A_\mu \psi. $$
We obtain equations which are the analogues of the abelian ones:
\begin{eqnarray*}
D_\nu \fup & =  & -g \bar{\psi} \gamma^\mu \psi, \\
(i \partial_\mu \gamma^\mu -m) \psi & = & -g A_\mu \gamma^\mu \psi.
\end{eqnarray*}
Classical analogues, called the Wong equations, exist but since in
applications particles are quantum, we shall not derive them here.

\subsection{Wu--Yang criterion}
\hspace*{\parindent}What about the dynamics of monopoles?  Let us go
back to the abelian theory first.  We know that in the presence of a
magnetic monopole, the potential $A_\mu$ has to be patched, say by
overlapping northern and southern hemispheres on any sphere
surrounding the monopole.  We immediately face several difficulties.
\begin{enumerate}
\item[(a)] Varying patched $A_\mu$;
\item[(b)] As the monopole moves, the patching moves with it, thus
depending on say $Y (\tau)$;
\item[(c)] What is the interaction term?
\end{enumerate}

\subsubsection*{Wu--Yang criterion:}
\hspace*{\parindent}Equations of motion are obtained by varying the
free field and free particle action, {\em under the constraint} that
there exists a magnetic monopole.  In other words, the interaction
comes from the constraint.  Intuitively, this is quite reasonable.
Around the monopole, the field configurations $A_\mu$ have to be such
that we have a nontrivial bundle.  As the monopole moves in spacetime,
the field configurations have to rearrange themselves to maintain this
topological condition, hence there is mutual influence, that is,
interaction.

So the prescription is:
$${\mathcal A}^0 = {\mathcal A}^0_F + {\mathcal A}_M^0 $$
to be varied under the constraint:
$$\partial_\nu \fstarup = -\tilde{\jmath}^\mu, $$
where we can insert either the classical or the quantum expression for
the current.  

Using the method of Lagrange multipliers, we in fact vary the
auxiliary action:
$$ {\mathcal A}' = {\mathcal A}^0 + \int \lambda_\mu (\partial_\nu
\fstarup + \tilde{\jmath}^\mu). $$

The next question is, what are the variables?   If we use $A_\mu$, we
still have the troubles (a) and (b).  To answer this, let us look at
the simpler problem of pure electromagnetism.

Previously, to get the free Maxwell equations we varied ${\mathcal
A}_F^0$ with respect to $A_\mu$.  Now we could  also have varied 
with respect to \fdown, {\em provided} we compensate for the fact that
there are more \fdown\ than $A_\mu$, the former with 6 degrees of
freedom and the latter with 4.
Suppose now we restrict ourselves only to those \fdown\ satisfying
$\partial_\nu \fstarup = 0$, then we shall be able to recover $A_\mu$
(at least in a local region).  This is clear in the language of forms,
because
$$ \partial_\nu \fstarup =0 \longleftrightarrow dF=0 $$
and in flat spacetime the Poincar\'e lemma will tell us that there
exist a 1-form $A$ such that $F=dA$.   So in other words, the sets of
variables $\{A_\mu (x)\}$ and $\{\fdown (x)\ {\rm with}\ \partial_\nu
\fstarup =0 \}$ are entirely equivalent.   So suppose we form the
action
$$ {\mathcal A}' = {\mathcal A}^0_F + \int \lambda_\mu (x)
(\partial_\nu \fstarup) $$
and vary with respect to \fdown, we shall indeed get from
$$ {\mathcal A}' = -\quart \int (\fdown \fup - \half
\epsilon^{\rho\sigma\mu\nu} \lambda_\rho \partial_\sigma \fdown), $$
giving
\begin{eqnarray*}
\fup &=& 2 \epsup \partial_\rho \lambda_\sigma \\
\Longrightarrow \partial_\nu \fup & = & 0,
\end{eqnarray*}
which is the desired Maxwell's equation.

So we see that the two derivations are entirely equivalent.

Coming back to the point monopole,  the constraint is
$$\partial_\nu \fstarup = - \tilde{e} \int d\tau \frac{dY^\mu}{d\tau}
\delta (x - Y(\tau)).  $$
We see that away from the monopole worldline, we have again
$$ \partial_\nu \fstarup = 0 \Longleftrightarrow \exists A_\mu. $$
Along the monopole worldline, we know already that no $A_\mu$ exists.
Hence in the constrained action principle, we are justified in using
\fdown\ as variables.  Hence the Wu--Yang criterion gives us the
dynamics as follows:
$$ \left\{ \begin{array}{rcll}
\partial_\nu \fup & = & 0 & {\rm no\ electric\ charges}\\
m \frac{d^2 Y^\mu (\tau)}{d\tau^2} & = & - \tilde{e} \fstarup
(Y(\tau)) \frac{dY_\nu (\tau)}{d\tau} & \\
\partial_\nu \fstarup & = & - \tilde{e} \int d\tau \frac{dY^\mu
(\tau)}{d\tau} \delta (x-Y(\tau)) & {\rm constraint}
\end{array} \right. $$
These are identical to the {\em dual} of the equations of motion of an
electrically charged particle, as expected---we shall study
electromagnetic duality in more details later on.

What about nonabelian theory?  In principle, the Wu--Yang criterion
can be applied to nonabelian monopoles.  In fact, it looks highly
plausible that it can be applied to any topological charges to find
their dynamics.  The difficulty in the case of the nonabelian monopole
is to write down the constraint.  {\em Recall} the charge is defined
as an element of $\pi_1 (G)$.

This programme has in fact been carried out using loop space
variables.  Since Yang--Mills theory is not symmetric under the Hodge
star ${}^*$ (as we shall show) the equations thus obtained are {\em
new}.   Unfortunately it is a bit too lengthy to present here.  What I
shall do is to indicate to you how to use the Wu--Yang criterion in
the pure Yang--Mills case to re-obtain the Yang--Mills equation, just
as I did above for the Maxwell case.

{\em Recall} the constraint for the existence of $A_\mu$, in terms of
loop variables is that the `connection'
$$ F_\mu (\xi,s) = \frac{i}{g}\;\Phi^{-1} (\xi)\;(\delta_\mu (s) \Phi
(\xi)) $$
is flat, that is, its `curvature' vanishes: 
$$ G_{\mu\nu}(\xi,s) = \delta_\nu (s) F_\mu(\xi,s) - \delta_\mu (s) 
F_\nu(\xi,s) +ig[F_\mu(\xi,s),F_\nu(\xi,s)] = 0. $$

Next we want to write the Yang--Mills action ${\mathcal A}^F_0$ in
loop variables.  It turns out that, modulo uncertainties about
measures in function spaces,
$${\mathcal A}^F_0 = - \frac{1}{\bar{N}} \int \delta \xi \int_0^{2\pi}
ds \Tr (F_\mu(\xi,s) F^\mu(\xi,s)) (\dot{\xi}^\alpha
\dot{\xi}_\alpha)^{-1}, $$
where $\bar{N}$ is an infinite normalization factor, the explicit
expression for which need not bother us here.

So by the Wu--Yang criterion we form the constrained action with the
Lagrange multipliers $L_{\mu\nu}$:
$$ {\mathcal A}' = {\mathcal A}_F^0 + \int \delta \xi \int ds \Tr
(L^{\mu\nu} (\xi,s) G_{\mu\nu} (\xi,s)).  $$
Because $G_{\mu\nu}$ is skew, we can choose without loss of generality
$L^{\mu\nu}$ to be skew as well.   Varying with respect to $F_\mu$ we
get
$$ F^\mu = - \bar{N} \dot{\xi}^2 (\delta_\nu L^{\mu\nu} -ig [F_\nu,
L^{\mu\nu}]).  $$
Write this as a loop covariant derivative ${\mathcal D}$
$$F^\mu = - \bar{N} \dot{\xi}^2 ({\mathcal D}_\nu L^{\mu\nu}). $$
Then 
\begin{eqnarray*}
{\mathcal D}_\mu F^\mu &=& - \bar{N} \dot{\xi}^2 ({\mathcal D}_\mu
{\mathcal D}_\nu L^{\mu\nu}) \\
&= & - \frac{\bar{N}}{2} \dot{\xi}^2 ([{\mathcal D}_\mu, {\mathcal
D}_\nu] L^{\mu\nu}) \quad (\because L^{\mu\nu}\ {\rm skew})\\
& = & 0 \qquad \qquad (\because F_\mu\ {\rm is\ flat}).
\end{eqnarray*}
But ${\mathcal D}_\mu F^\mu = \delta_\mu F^\mu - ig [F_\mu,F^\mu]$ and
the commutator term vanishes, from which we obtain
$$ \delta_\mu (s) F^\mu (\xi,s) = 0. $$
This we refer to as the Polyakov equation, and is shown by Polyakov
[AM Polyakov: Nucl.\ Phys.\ B164 (1980) 171--188] to be equivalent to
Yang--Mills equation:
$$D_\nu \fup =0. $$
So we see that again the Wu--Yang criterion gives us the right
dynamics.

\smallskip

\noindent{\bf Note.}\ \ We have used the Wu-Yang criterion quite
extensively, and have recovered all the known equations for
interaction between the gauge field and charges, and have also
obtained some new equations as well, notably for the nonabelian 
monopoles.

\subsection{Symmetry breaking}
\hspace*{\parindent}From Lecture 2 we learned that we need to consider
more complicated gauge theories, that is, those that exhibit symmetry
breaking as in electroweak theory.

We shall now look at the action for electroweak theory.  For
simplicity and to make the symmetry breaking mechanism more
transparent, we shall omit the charges (i.e.\ leptons).

To the Yang--Mills action ${\mathcal A}_F^0$ we now add the Higgs
action:
$${\mathcal A}_H = \int (D_\mu \phi D^\mu \phi + V(\phi)) d^4 x, $$
where $\phi$ is an $SU(2)$ doublet of  complex scalar fields
$$ \phi = \left( \begin{array}{c}
\phi^+ \\ \phi^0 \end{array} \right) $$
with weak hypercharge $Y=-\half$.  Hence the covariant derivative is
$$D_\mu \phi = (\partial_\mu - \tfrac{i}{2} g_2 
\mbox{\boldmath $\tau$} \cdot
\mathbf{W}_\mu + \tfrac{i}{2} g_1 Y_\mu) \phi, $$
where $\mathbf{W}_\mu$ represents the 3 components of the $SU(2)$ weak
isospin gauge potential, and $Y_\mu$ the $U(1)$ weak hypercharge
potential, and $g_2$ and $g_1$ the corresponding coupling constants.
The potential is
$$ V(\phi) = -\tfrac{\mu^2}{2} |\phi|^2 - \tfrac{\lambda}{4}
|\phi|^4 \quad (\lambda >0). $$
If $\mu^2 >0$, then the scalar field $\phi$ has mass $\mu$ and the
vacuum (or ground state) corresponds to $\phi_0 =0$.  If $\mu^2 <0$,
we get the famous Mexican hat potential, and the vacuum (with
$V(\phi)$ minimum) is given by
$$ |\phi_0| = - \mu^2/\lambda =\eta \ne 0. $$
We now choose a gauge such that
$$ \phi_0= \eta \left( \begin{array}{c}
0\\1 \end{array} \right). $$
In this way, the vacuum corresponds to a particular direction in the
space of $\mathfrak{su} (2) \oplus \mathfrak{u} (1)$ and once this 
choice is
made, the physics will no longer be invariant under the whole of the
$U(2)$ group.  This is {\em symmetry breaking}.  In fact, since 
$\phi$ is a complex vector in $\bbc^2$ (although we sometimes call it
a spinor), there will be a phase rotation left over after fixing a
directin as above, and this constitutes the `little group' $U(1)$
corresponding to electromagnetism. 

For a quantum field theory, we look at quantum excitations around the
vacuum, that is,
$$ \phi(x) = \left( \begin{array}{c}
0 \\ \eta + \tfrac{\sigma (x)}{\sqrt{2}} \end{array} \right), $$
where $\sigma (x) \in \bbr$, which is a gauge choice.  Hence
$$D_\mu \phi = \left( \begin{array}{c} 
0 \\ \tfrac{1}{\sqrt{2}} \partial_\mu \sigma \end{array} \right) -
\left( \frac{ig_2}{2} \left( \begin{array}{cc}
W_\mu^3 & W_\mu^1 -i W_\mu^2 \\ W_\mu^1 + i W_\mu^2 & -W_\mu^3
\end{array} \right) - \frac{ig_1}{2} Y_\mu \right) 
\left( \begin{array}{c}
0 \\ \eta + \tfrac{\sigma (x)}{\sqrt{2}} \end{array} \right), $$
from which
$$ D_\mu \phi D^\mu \phi = \half (\partial_\mu \sigma)^2 +
\tfrac{g_2^2 \eta^2}{4} ((W_\mu^1)^2 + (W_\mu^2)^2) +
\tfrac{\eta^2}{4} (g_2 W_\mu^3 + g_1 Y_\mu)^2 + {\rm cubic} + {\rm
quartic}. $$
Now define: 
\begin{eqnarray*}
A_\mu & = & -\frac{e}{g_2} W_\mu^3 + \frac{e}{g_1} Y_\mu \\
Z_\mu & = & \frac{e}{g_1} W_\mu^3 + \frac{e}{g_2} Y_\mu
\end{eqnarray*}
with 
$$e=\frac{g_1g_2}{\sqrt{g_1^2 + g_2^2}}; $$
or in terms of the Weinberg angle
$$ \sin \theta_W = \frac{g_1}{\sqrt{g_1^2 + g_2^2}}$$
we have
\begin{eqnarray*}
A_\mu & = & -\sin \theta_W\,W_\mu^3 + \cos \theta_W\,Y_\mu \\
Z_\mu & = & \cos \theta_W\,W_\mu^3 + \sin \theta_W\,Y_\mu.
\end{eqnarray*}

As far as the particle spectrum is concerned, cubic and quartic terms
are unimportant.  They can be either got rid of by redefining fields or
they represent self-interactions.  So we concentrate on the quadratic
terms.

{\em Recall} the Klein--Gordon Lagrangian
$$-(\partial_\mu \phi \partial^\mu \phi + m^2 \phi)  \leadsto
 \phi (\partial_\mu \partial^\mu - m^2) \phi $$
(after integration by parts).  So from the previous expression for
$D_\mu \phi D^\mu \phi$ plus the Yang--Mills action, we see that the
following fields acquire a nonzero mass term, giving
$$M^2_{W^1} = M^2_{W^2} = M^2_W = \frac{g_2^2 \eta^2}{2}, \quad M^2_Z =
\frac{g^2_2 \eta^2}{2 \cos^2 \theta_W} = \frac{M^2_W}{\cos^2
\theta_W}, $$
(and also the Higgs field $\sigma$ becomes massive) while the abelian
vector field
$$A_\mu =-\sin \theta_W\,W_\mu^3 + \cos \theta_W\,Y_\mu $$
has no mass term and hence remains massless.  This can easily be
identified as the photon (especially if we consider the lepton terms
as well). 

\clearpage

\section{Electric--magnetic duality}
\hspace*{\parindent}Electric--magnetic duality, where it exists, is an important concept in
theoretical physics.
\begin{enumerate}
\item As a symmetry of nature, we should study it.  Also since it is
discrete, it should be relatively easy.
\item As a result of this symmetry we need study only half of the 
phenomena.
\item Dirac's quantization condition (abelian and nonabelian
respectively) says
$$ e \tilde{e}= 2\pi, \quad g \tilde{g}=4 \pi. $$
This should hold even under renormalization.  We have therefore a
correspondence which relates weak coupling (where perturbation
expansion is good) to strong coupling (where perturbation expansion is
bad).
\item 't~Hooft's theorem (see later) leads to a mechanism for
confinement (of quarks) via duality.
\end{enumerate}

\subsection{Abelian theory}
\hspace*{\parindent}We recall that the duality operator 
(${}^*$) is defined by:
$$ {}^*\!F_{\mu\nu} = -\half \epsilon_{\mu\nu\rho\sigma}
F^{\rho\sigma},$$
the sign being the consequence of Minkowski signature $(+---)$.  

Duality, as the name implies, is such that we recover the original
field tensor (up to sign) if we do the operation twice:
$$ {}^* ({}^*\!F) = - F. $$

In terms of the electric field $E$ and the magnetic field $B$ these
tensors can be represented in matrix form:
$$F_{\mu\nu}= \left( \begin{array}{rrrr}
0&E_1&E_2&E_3\\
-E_1&0&-B_3&B_2\\
-E_2&B_3&0&-B_1\\
-E_3&-B_2&B_1&0
\end{array} \right) \quad {}^*\!F_{\mu\nu}= \left( \begin{array}{rrrr}
0&B_1&B_2&B_3\\
-B_1&0&E_3&-E_2\\
-B_2&-E_3&0&E_1\\
-B_3&E_2&-E_1&0
\end{array} \right). $$

So we see that under $\ {}^* \colon \mathbf{E} \to \mathbf{B},\ \
\mathbf{B} \to -\mathbf{E}.$  

The question is: is the theory {\em invariant} under this {\em
duality}?

In vacuo, the Maxwell equations are:
\begin{eqnarray*}
(1)\quad \partial_\nu F^{\mu\nu}&=&0,\\
(2)\quad \partial_\nu {}^*\!F^{\mu\nu}&=&0.
\end{eqnarray*}
Hence we conclude that the theory is indeed invariant in this case.

But we can go back even further and see that the derivation of these
equations is also invariant.  In this connection we note that
$$ {\mathcal A}^0_F = -\quart \int F_{\mu\nu} F^{\mu\nu} = \quart
\int {}^*\!F_{\mu\nu} {}^*\!F^{\mu\nu}, $$
so that the action is invariant (the $-$ sign being of no significance
as it does not affect the dynamics).  Applying the Wu--Yang criterion,
we can use either (1) or (2) as constraint and obtain the other as
equation of motion, so that we end up with the same equations in both
cases.

Recall that
$$ {\mathcal A}' = -\quart \int (F_{\mu\nu} F^{\mu\nu} - \half
\epsilon_{\mu\nu\rho\sigma} \lambda_\rho \partial_\sigma F_{\mu\nu}),
$$
leading to
$$ F^{\mu\nu}= 2 \epsilon^{\mu\nu\rho\sigma} \partial_\nu \lambda_\mu,
$$
which implies
\begin{eqnarray*}
\partial_\nu F^{\mu\nu} &=&0\\
{}^*\!F_{\mu\nu}&=& \partial_\nu \tilde{A}_\mu - \partial_\mu
\tilde{A}_\nu, \ \ \tilde{A}_\mu=\lambda_\mu.
\end{eqnarray*}

\vspace*{5mm}

This is clear by looking at Chart \ref{chart1}, both columns 2 and 3.  

Next, in the 
presence of a magnetic monopole, we have:
\begin{enumerate}
\item[(1)] $\partial_\nu F^{\mu\nu}=0$
\item[(2)] $\partial_\nu {}^*\!F^{\mu\nu} = -\tilde{\jmath}^\mu$.
\end{enumerate}

If we now look at Chart \ref{chart2}, column 1 or 2, we see that:
\begin{center}
(2) as constraint $\leadsto$ (1) as equation of motion.
\end{center}

Dually, in the presence of an electric  charge, we have
\begin{enumerate}
\item[(1)] $ \partial_\nu F^{\mu\nu}=-j^\mu$
\item[(2)] $\partial_\nu {}^*\!F_{\mu\nu} = 0$.
\end{enumerate}

And now Chart \ref{chart2}, column 3 or 4, tells us:
\begin{center}
(1) as constraint $\leadsto$ (2) as equation of motion.
\end{center}

\noindent{\bf Remarks}
\begin{enumerate}
\item Duality: $E \leftrightarrow B, \ e \leftrightarrow \tilde{e},\
j^\mu \leftrightarrow \tilde{\jmath}^\mu$.  In other words, the duality 
links charges $ \leftrightarrow$ monopoles, and which is which depends on 
which are the fields/potentials.
\item A dual potential $\tilde{A}_\mu$ emerges, which is just the
Lagrange multiplier.  Away from charges and monopoles, we have both
$A_\mu$ and $\tilde{A}_\mu$.  In pure theory, neither $A_\mu$ nor 
$\tilde{A}_\mu$ appears in the equations, but in the presence of
charges/monopoles, the Lagrange multiplers cannot be eliminated and
the potentials appear explicitly, as {\em demanded} by the
Bohm--Aharonov experiment.
\item The duality goes deeper, as linking physics with geometry.
$$
\begin{array}{ccc}
\fbox{\shortstack{$A_\mu$ exists as \\ potential for $F_{\mu\nu}$ \\ 
$(F=dA)$}} & 
\stackrel{{\rm Poicar\acute{e}}}{\Longleftrightarrow}
& \fbox{\shortstack{$\partial_\mu \fstarup=0$ \\ $(dF =0)$}} \\
&&\\
\Big\Updownarrow & & \Big\Updownarrow\vcenter{%
\rlap{$\scriptstyle{\rm Gauss}$}}\\
&& \\
\fbox{\shortstack{Principal $A_\mu$ \\ bundle trivial}}  & 
 &
\fbox{\shortstack{No magnetic\\ monopole $\tilde{e}$}}\\
&&\\
{\rm GEOMETRY} && {\rm PHYSICS}
\end{array}
$$
Dually, we have exactly the same picture:
$$
\begin{array}{ccc}
\fbox{\shortstack{$\tilde{A}_\mu$ exists as \\ potential for 
${}^*\!F_{\mu\nu}$ \\ 
$({}^*\!F=d\tilde{A})$}} & 
\stackrel{{\rm Poicar\acute{e}}}{\Longleftrightarrow}
& \fbox{\shortstack{$\partial_\mu F^{\mu\nu}=0$ \\ $(d\,{}^*\!F =0)$}} \\
&&\\
\Big\Updownarrow & & \Big\Updownarrow\vcenter{%
\rlap{$\scriptstyle{\rm Gauss}$}}\\
&& \\
\fbox{\shortstack{Principal $\tilde{A}_\mu$ \\ bundle trivial}}  & 
 &
\fbox{\shortstack{No electric \\ charge $e$}}\\
&&\\
{\rm GEOMETRY} && {\rm PHYSICS}
\end{array}
$$
\end{enumerate}

\subsection{The star operation in Yang--Mills theory}
In nonabelian theory, we have the same star operation:
$$ {}^*\!F_{\mu\nu} = -\half \epsilon_{\mu\nu\rho\sigma}
F^{\rho\sigma}.$$

Let us now try to fill in the boxes as in the abelian case.  We find
in the direct picture:
$$
\begin{array}{ccc}
\fbox{\shortstack{$A_\mu$ exists as \\ potential for $F_{\mu\nu}$ \\ 
$(F=D_A A)$}} & 
\stackrel{{\rm Bianchi}}{\Longrightarrow}
& \fbox{\shortstack{$D_\mu \fstarup=0$ \\ $(D_A F =0)$}} \\
&&\\
\Big\Updownarrow & & \Big\vert\vcenter{%
\rlap{$\scriptstyle{?}$}}\\
&& \\
\fbox{\shortstack{Principal $A_\mu$ \\ bundle trivial}}  & 
\stackrel{{\rm definition}}{\Longleftrightarrow} &
\fbox{\shortstack{No magnetic\\ monopole $\tilde{g}$}}
\end{array}
$$

And in the dual picture:
$$
\begin{array}{ccc}
\fbox{\shortstack{$\tilde{A}_\mu$ exists as\\ potential for 
${}^*\!F_{\mu\nu}$ 
\\ $({}^*\!F=D_{\tilde{A}} \tilde{A})$}} & 
\stackrel{{\rm Gu-Yang}}{\not\!\!\!\Longleftarrow}
& \fbox{\shortstack{$D_\mu F^{\mu\nu}=0$ \\ $(D_A {}^*\!F =0)$}} \\
&&\\
\Big\Updownarrow & & \Big\Updownarrow\vcenter{%
\rlap{$\scriptstyle{\rm YM}$}}\\
&& \\
\fbox{\shortstack{Principal $\tilde{A}_\mu$ \\ bundle trivial}}  & 
\stackrel{?}{\cdots\cdots} &
\fbox{\shortstack{No electric\\ charge $g$}}
\end{array}
$$

All these go to show that the star operation in nonabelian theory does
not give us the desired electric--magnetic duality, unlike the abelian
case.  In fact, we have a stronger result, in the following
counter-example discovered by Gu and Yang.

\vspace*{5mm}

\noindent{\bf The Gu--Yang counter-example}

\medskip

Gu and Yang phrase their example in terms of $D_A {}^*\!F$, but we can 
equally think in terms of $D_A F$.  Since in general $D_A F =0\  
\not\!\!\Longrightarrow dF=0$, there is really no reason to suppose
that $F$ is 
in any sense exact.  Furthermore, we are {\em not} asking $F$ to be exact, 
but we want the existence of $A$ for which $F=D_A A$.  So the existence
of a {\em gauge potential} in nonabelian theory has very little to do
with the usual Poincar\'e lemma.

\medskip

{\bf Counter-example.}\ \ Let $G=SU(2)$.  Take an explicit `hedgehog
potential'
$$A_i^j= \epsilon_{ijk} x_k g(r), \quad A_0^j =0, $$
(abandoning our summation convention temporarily) with $g$ a function 
of the radius $r$ satisfying
$$ (gr)'' + \frac{2}{r} (gr)' - (1+r^2g)(\frac{2g}{r} +rg^2)=0. $$  

It can easily be verified that for such a potential the gauge field
satisfies the Yang--Mills source-free equation:
$$ D_\nu F^{\mu\nu} =0. $$

Previously, Wu and Yang found numerical solutions to the above differential
equation for $g$, depending on a real parameter $c>0$, with
\begin{eqnarray*}
g(r) & = & -\frac{c}{r^2} + O(\frac{1}{r^3}) \quad {\rm as}\ r \to
\infty \\
g(r) & \to & -\frac{1}{r} \quad {\rm as}\ r \to 0.
\end{eqnarray*}

Now let 
$$V_1^j = F_{23}^j, \quad V_2^j=F_{31}^j, \quad V_3^j=F_{12}^j, $$
and consider $\vec{V}_1, \vec{V}_2, \vec{V}_3$ as 3 vectors in
3-space.  Then it can be proved that
$$ \vec{V}_1, \vec{V}_2, \vec{V}_3 \ {\rm linearly\ indept}
\Leftrightarrow g(r) \ne \frac{a}{r^2}\ {\rm for\ any}\ a.  $$
Hence for the given solution, the vectors are linearly independent.

Suppose for a contradiction that $\tilde{A}_\mu$ exists as a potential
for \fstarup.  Then its Bianchi identity is
$$ \partial_\mu F^{\mu\nu} - i \tilde{g} [\tilde{A}_\mu, F^{\mu\nu}]
=0. $$
This, together with the Yang--Mills source-free equation, implies
$$[A_\mu - \tilde{A}_\mu, F^{\mu\nu}] =0, $$
where for convenience we have absorbed the coupling constant
$\tilde{g}$ into  $\tilde{A}_\mu$, $g$ into $A_\mu$.
Notice that $\tilde{A}_0$ may not be zero, but it does not contribute
because $F^{0\nu}=0$.

Write now $U^j_k =A^j_k - \tilde{A}^j_k$.  Then in 3-space notation,
the commutator equation can be written as:
\begin{eqnarray*}
\vec{U}_2 \wedge \vec{V}_3  -  \vec{U}_3 \wedge \vec{V}_2 & = & 0,
\\
\vec{U}_1 \wedge \vec{V}_3 - \vec{U}_3 \wedge \vec{V}_1&=&0,\\
\vec{U}_1 \wedge \vec{V}_2 - \vec{U}_2 \wedge \vec{V}_1&=&0.
\end{eqnarray*}

We now claim that: $\vec{V}_i\ {\rm lin.\ indept.}\ \Longrightarrow
\vec{U}_i =0. $

In fact, all the quadruples $(\vec{U}_2, \vec{U}_3, \vec{V}_2,
\vec{V}_3 )$ etc.\ are coplanar.  This implies, say, 
\begin{eqnarray*}
\vec{U}_2 + \vec{U}_3 & = & \alpha_2 \vec{V}_2 + \alpha_3
\vec{V}_3 \\
\vec{U}_1 + \vec{U}_2 & = & \beta_1 \vec{V}_3 + \beta_2 
\vec{V}_2,
\end{eqnarray*}
which in turn implies
$$\alpha_2=\beta_2, \quad {\rm etc.}  $$
Hence
\begin{eqnarray*}
\vec{U}_1 + \vec{U}_2 & = & \alpha_1 \vec{V}_1 + \alpha_2
\vec{V}_2, \\
\vec{U}_2 + \vec{U}_3 & = & \alpha_2 \vec{V}_2 + \alpha_2
\vec{V}_2, \\
\vec{U}_1 + \vec{U}_3 & = & \alpha_1 \vec{V}_1 + \alpha_3
\vec{V}_3,
\end{eqnarray*}
and therefore all $\alpha_i$ are equal, say to $\alpha$.  This gives
$$ 
\begin{array}{crcl}
&\vec{U}_i & =& \alpha \vec{V}_i,\  \forall i\\
\Longrightarrow & \alpha(\vec{V}_2 \wedge \vec{V}_3) &=& \alpha 
(\vec{V}_3 \wedge \vec{V}_2)\\
\Longrightarrow & \alpha & = & 0,
\end{array} $$
which justifies our claim.

We therefore conclude:
$$ \tilde{A}_k^j = A_k^j, \quad \forall k,j. $$
Now
$$ {}^*\!F_{32} =F_{10} =0, $$
but on the other hand
$$ {}^*\!F_{32} =F_{32} = -\vec{V}_1 \ne 0, $$
which is a contradiction.

{\bf Claim.}\quad Hence nonabelian Yang--Mills theory is not dual symmetric 
under ${}^*$.
\begin{enumerate}
\item $\tilde{A}_\mu$ need not exist,
\item the dual of Yang-Mills equation is {\em not} Bianchi identity.
\end{enumerate}

\subsection{Generalized electric--magnetic dualtiy for Yang--Mills
theory}
We saw that electromagnetic duality in Maxwell theory is both
important and useful.   So we want to salvage the situation as regards
to nonabelian theory.   There are two ways of going about it.
\begin{enumerate}
\item[(A)] Generalize the concept of duality, that is, modify it in
the nonabelian case.
\item[(B)] Enrich the theory, for example, make it supersymmetric, so
as to enlarge existing symmetries.
\end{enumerate}

We shall briefly talk about both.  Take (A) first.  We seek a dual
transform $(\tilde{\ })$ satisfying the following properties:
\begin{enumerate}
\item $(\quad)^{\sim\sim} = \pm (\quad)$,
\item electric field $F_{\mu\nu} \stackrel{\sim}{\longleftrightarrow}
$ magnetic fields $\tilde{F}_{\mu\nu}$,
\item both $A_\mu$ and $\tilde{A}_\mu$ exist as potentials (away from
charges),
\item magnetic charges are monopoles of $A_\mu$, and electric charges
are monopoles of $\tilde{A}_\mu$,
\item $\tilde{\ }$ reduces to ${}^*$ in the abelian case.
\end{enumerate}

So far, we are only able to express this new dual transform in terms
of loop variables.  I cannot go through the construction here but
those interested can refer to our papers, especially recent reviews.
Suffice it to say that the above 5 points are indeed satisfied, and we
have full duality as depicted below:
$$
\begin{array}{ccc}
\fbox{$A_\mu$ exists} & 
\stackrel{{\rm extended\ Poicar\acute{e}}}{\Longleftrightarrow}
& \fbox{\shortstack{loop space formula \\ $\tilde{D}_\nu
\tilde{F}^{\mu\nu}=0\ ({\rm YM})^\sim$}} \\
&&\\
\Big\Updownarrow & & \Big\Updownarrow\\
&& \\
\fbox{\shortstack{Principal $A_\mu$ \\ bundle trivial}}  & 
\stackrel{\rm definition}{\Longleftrightarrow}
 &
\fbox{No magnetic charge}\\
&&\\
{\rm GEOMETRY} && {\rm PHYSICS}
\end{array}
$$
Dually, we have
$$
\begin{array}{ccc}
\fbox{$\tilde{A}_\mu$ exists} & 
\stackrel{{\rm extended\ Poicar\acute{e}}}{\Longleftrightarrow}
& \fbox{\shortstack{(loop space formula)$^\sim$ \\ $D_\mu
F^{\mu\nu}=0\ ({\rm YM})$}} \\
&&\\
\Big\Updownarrow & & \Big\Updownarrow\vcenter{%
\rlap{$\scriptstyle{\rm definition}$}}\\
&& \\
\fbox{\shortstack{Principal $A_\mu$ \\ bundle trivial}}  & 
\Longleftrightarrow &
\fbox{No electric charge}\\
&&\\
{\rm GEOMETRY} && {\rm PHYSICS}
\end{array}
$$

\subsection{'t Hooft's theorem and its consequences}
In a quantum gauge field theory, the phase factor
$$ \Phi (C)=P_s \exp ig \int_C A_\mu (x) dx^\mu $$
is an operator in a Hilbert space.  Let
$$ A(C) = \tr \Phi (C), $$
which is still an operator.  This is the more usual definition of the
{\em Wilson loop}.

In a gauge theory with gauge group corresponding to the Lie algebra 
$\mathfrak{su}(N)$, 't~Hooft introduced abstractly the operator
$B(C')$ {\em dual} to $A(C)$, by the commutation relation
$$ A(C) B(C') = B(C') A(C) \exp (2\pi i n/N), $$
where $n$ is the linking number between the two {\em spatial} loops
$C$ and $C'$.  He describes these two quantities as:
\begin{itemize}
\item $A(C)$ measures the magnetic flux through $C$ and creates
electric flux along $C$
\item $B(C)$ measures the electric flux through $C$ and creates
magnetic flux along $C$
\end{itemize}

So they play dual roles in the sense we have been considering.
However, there was no ``magnetic'' potential available at the time, so
that the definition of $B(C')$ was not explicit, only through the
commutation relation above.

But in the construction mentioned in the last subsection (which I did
not give explicitly), the magnetic potential $\tilde{A}_\mu$ exists, 
so that one can actually prove the commutation
relation.  This has been done about 2 years ago.

\vspace*{3mm}

{\noindent}{\bf 't Hooft's Theorem.}\ \ {\em If the Wilson loop
operator of an $SU(N)$ theory and its dual theory satisfy the
commutation relation given above, then:}
\begin{eqnarray*}
SU(N)\ {\rm confined} &\Longleftrightarrow & \widetilde{SU(N)}\ 
{\rm broken}  \\
SU(N)\ {\rm broken} &\Longleftrightarrow & \widetilde{SU(N)}\ 
{\rm confined}
\end{eqnarray*}

\vspace*{3mm}

Note that the second statement follows from the first, given that the
operation of duality is its own inverse (up to sign).

The theorem does not hold for a $U(1)$ theory, where both $U(1)$ and
$\widetilde{U(1)}$ may exist in a Coulomb phase, that is, with long
range potential ($\sim 1/r$).

The statement is phrased in terms of phase transition, and has
profound implications.  It has been a cornerstone for attempts to
prove quark confinement ever since.

I may add that we have exploited 't~Hooft's theorem in the reverse
way.  Given that $SU(3)$ colour is confined, we deduce that dual
colour is broken, which we have identified as the 3 generations of
particles as observed in nature.  There are many consequences of such
a hypothesis, not only in particle physics, but also in nuclear and 
astrophysics.

\smallskip

Coming back to the commutation relation, I wish to show you how to
prove it in the abelian case, just to give you a taste of what is
involved.  The nonabelian case is too complicated to treat here.

In the abelian case, we do not need the trace, hence $A(C)=\Phi(C),\
B(C')= \tilde{\Phi}(C')$, and the $\Phi$ are genuine exponentials.  So
if we can show the following relation for the exponents, we shall have
proved the required commutation relation:
$$\left[ ie \oint_C A_i dx^i,\ i \tilde{e} \oint_{C'} \tilde{A}_i dx^i
\right] = 2\pi ni. $$

Using Stokes' theorem the second integral
$$=-i\tilde{e} \int\!\!\int_{\Sigma_{C'}} {}^*\!F_{ij} d \sigma^{ij} = 
 i\tilde{e} \int\!\!\int_{\Sigma_{C'}} E_i d\sigma^i,\ {\rm where}\ 
\partial\Sigma_{C'}=C'. $$

For simplicity, suppose the linking number $n=1$.  Then the loop $C$
will intersect $\Sigma_{C'}$ at some point $x_0$---if it intersects
more than once, the other contributions will cancel in pairs, so we
shall ignore them.  So except for $x_0$, all points in $C$ are
spatially separated from points on $\Sigma_{C'}$.

Using the canonical commutation relation for $A_i$ and $E_j$
$$ [E_i (x), A_j (x')] = i \delta_{ij} \delta (x-x') $$
we get
$$\left[ ie \oint_C A_i dx^i,\ i\tilde{e} \int\!\!\int_{\Sigma_{C'}}
E_j d\sigma^j \right] = i e \tilde{e} = 2\pi i$$
by Dirac's quantization condition.

Hence we have shown explicitly in the abelian case that our definition
of duality coincides with 't~Hooft's.  The same is true in the
nonabelian case.

\subsection{Magnetic monopoles from symmetry breaking}
We looked at electroweak symmetry breaking in detail: $U(2) \to
U(1)$.  We can also do similar breaking with $SU(2) \to U(1)$.  Again
we can choose the Higgs field in the $\left( \begin{array}{c} 
0 \\ 1 \end{array} \right)$ direction, and the residual symmetry will
again be a $U(1)$, this time generated by the generator $T_3$.

Now $SU(2)$ being simply connected, there are no nontrivial bundles
over $S^2$.  However, there can be nontrivial reductions to the $U(1)$
subgroup.  Topologically, this can be seen by looking at part of the
exact sequence of homotopy groups, obtained from the principal bundle: 
$$ U(1) \to SU(2) \to SU(2)/U(1) \simeq S^2,  $$
whence
$$ \begin{array}{ccccccccc}
\to &\pi_2 (SU(2)) &\to& \pi_2 (SU(2)/U(1)) &
\stackrel{\sim}{\to} & \pi_1(U(1)) & \to & \pi_1 (SU(2)) & \to
\\
& \| &&&&&& \| & \\
& 0  &&&&&& 0  &
\end{array}  $$

The boundary condition of the Higgs field $\phi$ at infinity will
determine the nature of the reduced bundle (more precisely, its first
Chern class), that is, the homotopy class of the map: $S^2 \to
SU(2)/U(1) \simeq S^2$, the first $S^2$ being the sphere at infinity.
This is precisely given by $\pi_2 (SU(2)/U(1)) \simeq \bbz$, which by
the exactness of the above, is isomorphic to the magnetic charges of
the residual $U(1)$, namely $\pi_1 (U(1))$. 

Let us look at an example, the residual charge 1 magnetic monopole.  
It is a particular solution of the Yang--Mills--Higgs equations we 
saw before.

Inserting the asymptotic condition we get a solution, for large $r$, 
similar to the Wu--Yang potential we had:
$$ F^3_{0i} =0, \ F^3_{ij} = - \frac{1}{e r^3} \epsilon_{ijk} r^k,\
{\rm all\ others} =0 $$
$$ \Longrightarrow B_k = \frac{r^k}{e r^3}, $$
that is, a magnetic field in radial direction at infinity, which is
why this is referred to by Polyakov as the ``hedgehog solution''.
Such a solution is called a {\em 't~Hooft--Polyakov monopole}.

In the (unphysical) limit studied by Prasad and Summerfield,
$$   \left.
\begin{array}{ccc}
|\phi| & \to& 1 \\ \lambda & \to &0 \end{array} \right\} \  {\rm as}\ r
 \to \infty,
 $$
exact solutions exist for $\lambda=0=\mu$.  These are finite
energy solitonic solutions. 

There is a parameter occurring in the behaviour of $|\phi| \sim 1 -
M/r$, which represents the mass of the soliton, and it satisfies an
inequality given by Bogomolny.  If this bound is saturated, then we
have what is known as a BPS monopole.  These have been much studied
because they can be obtained by a process called {\em dimensional 
reduction} from instanton solutions of 4-dimensional Euclidean
self-dual Yang--Mills equations, as the gauge fields being static
have only spatial components, leaving the (imaginary) temporal
component of the instanton to play the role of the Higgs field.

Also these have been extended to exhibit electric charges as well.
The resulting extended `particle' is then a {\rm dyon}: carrying both
electric and magnetic charges.

\subsection{Seiberg--Witten duality}
The second way to study electric-magneitc duality is to exploit the
duality between the electric and magnetic charges which occurs as a
result of symmetry breaking from a nonabelian Yang--Mills theory.

In the models so far studied, supersymmetry is a necessary
ingredient.  Supersummetry is a hypothetical symmetry between fermions
and bosons, and has tremendous theoretical and mathematical appeal to
a lot of physicists.  We cannot discuss it here for lack of time (and
expertise!).   The dual symmetry I shall outline below works for all
$N=4$ and some $N=2$ supersymmetric Yang--Mills theories, with gauge
group $SU(2)$---this can be generalized.

The bosonic part of the action is
$${\cal A} = - \int \frac{1}{4} \Tr F_{\mu\nu} F^{\mu\nu} -
\frac{1}{2} \Tr D_\mu \phi D^\mu \phi - V(\phi) + \frac{\theta}{16
\pi^2} \Tr F_{\mu\nu} \fstarup, $$
where the last term corresponds to the second Chern class or instanton
number.  This is a topological term, and the coefficient in front of
it is the so-called $\theta$-vacuum angle.  Experimentally it is very
nearly 0.

By giving a nonzero vacuum expectation value $\eta$ to the Higgs field
$\phi$, we effect the symmetry breaking $SU(2) \to U(1)$.  There are
solutions which are BPS monopoles.  In fact they are dyons, with both
electric and magnetic charges.  Their masses satisfy the Bogomolny
bound:
$$ M^2 = (4\pi)^2 (m,n) \frac{\eta^2}{\Im \tau} \left( 
\begin{array}{cc}
1 & \Re \tau \\ \Re \tau & |\tau|^2 \end{array} \right) \left(
\begin{array}{c}
m \\ n \end{array} \right), $$
where 
$$ \tau = \frac{\theta}{2\pi} + i \frac{4\pi}{e^2},
\quad Q_m=\frac{n}{e}, \quad Q_e=e(m +  \frac{n \theta}{2\pi} ). $$
One sees that the mass formula is invariant under
$$ \tau \mapsto \frac{a \tau +b}{c \tau + d}, \quad \left( \begin{array}{c}
m \\ n \end{array} \right) \mapsto \left( \begin{array}{cc}
a & b \\ c& d \end{array} \right) \left( \begin{array}{c}
m \\ n \end{array} \right), $$
so that this theory of charges and monopoles are invariant under the
group $SL(2,\bbz)$.

In the particular case when $\theta=0$, the generator $S$ of
$SL(2,\bbz)$ corresponding to $a=d,\ b=c=-1$, that is $\tau \mapsto
-1/\tau$, induces
$$ \frac{e^2}{4\pi} \mapsto \frac{4\pi}{e^2}=\frac{\tilde{e}^2}{4\pi},
\ n \mapsto m, \ m \mapsto -n$$ 
and we recover the usual electric-magnetic dualtiy with $e \tilde{e} =
4\pi$.  This also goes under the name of {\em $S$-duality}.

Explicit solutions are constructed by making use of a certain
holomorphic function of $\tau$ occurring in the theory having to do
with the metric on the moduli space.  This duality is found to be a
symmetry of the quantum field theory.

Seibery and Witten also considered supersymmetric Yang--Mills theories
in which the dual symmetry is only partial, in the sense that the
spectrum of dyons in one theory is found to match the dual spectrum
(electric $\leftrightarrow$ magnetic) of another theory, perhaps with a
different gauge group.

The whole subject has been intensely studied in recent years, with
many ramifications into string theory, membrane theory and
$M$-theory.  They are definitely outside the scope of these lectures.

\clearpage

\begin{chart}
\vspace*{-2.5cm}
  \centerline{
    \resizebox{!}{23cm}
{\includegraphics{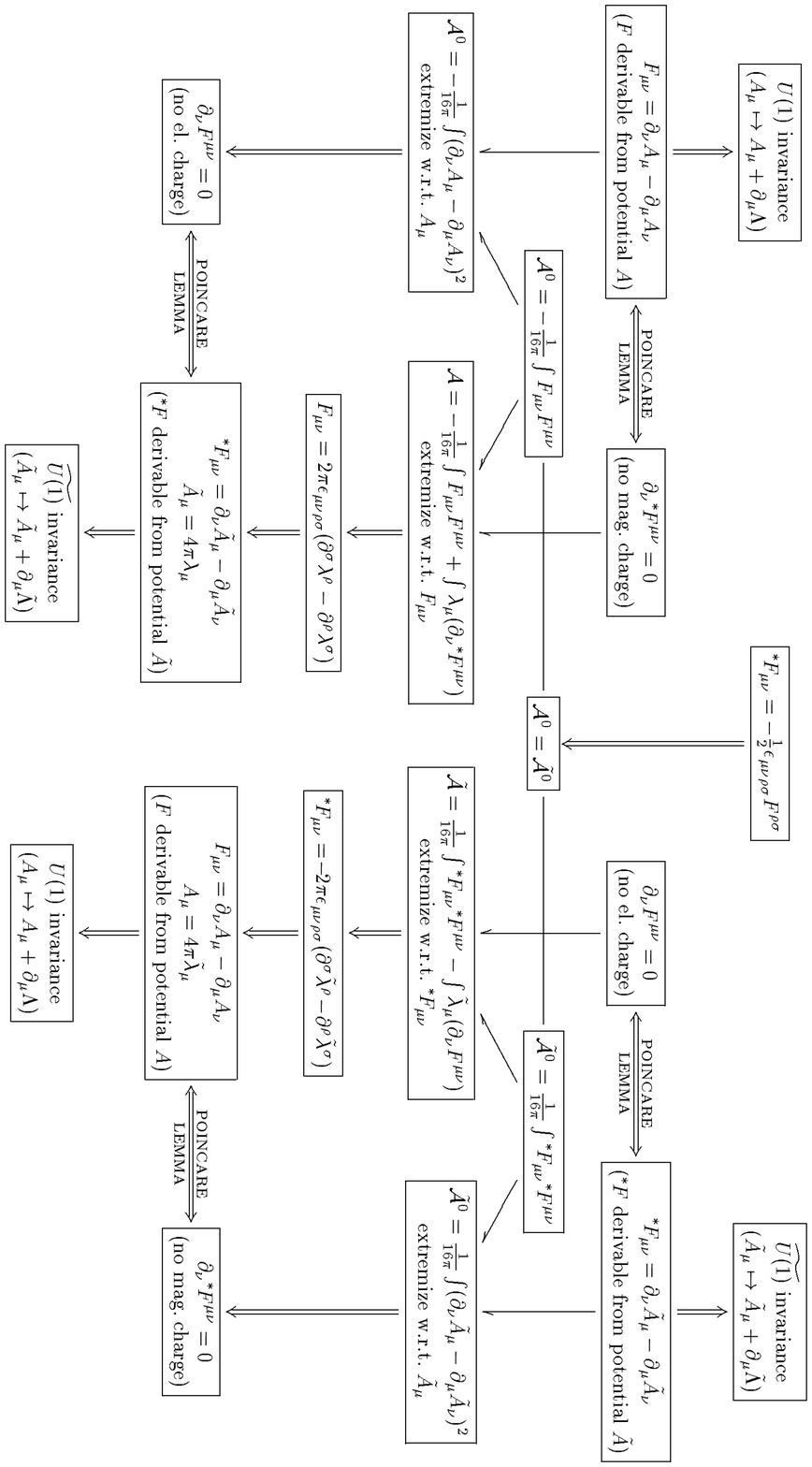}}
    }
\vspace*{-2.2cm}
 \caption[]{Pure Electromagnetism.}
  \label{chart1}
\end{chart}

\clearpage

\begin{chart}
\vspace*{-2.5cm}
  \centerline{
    \resizebox{!}{23cm}
{\includegraphics{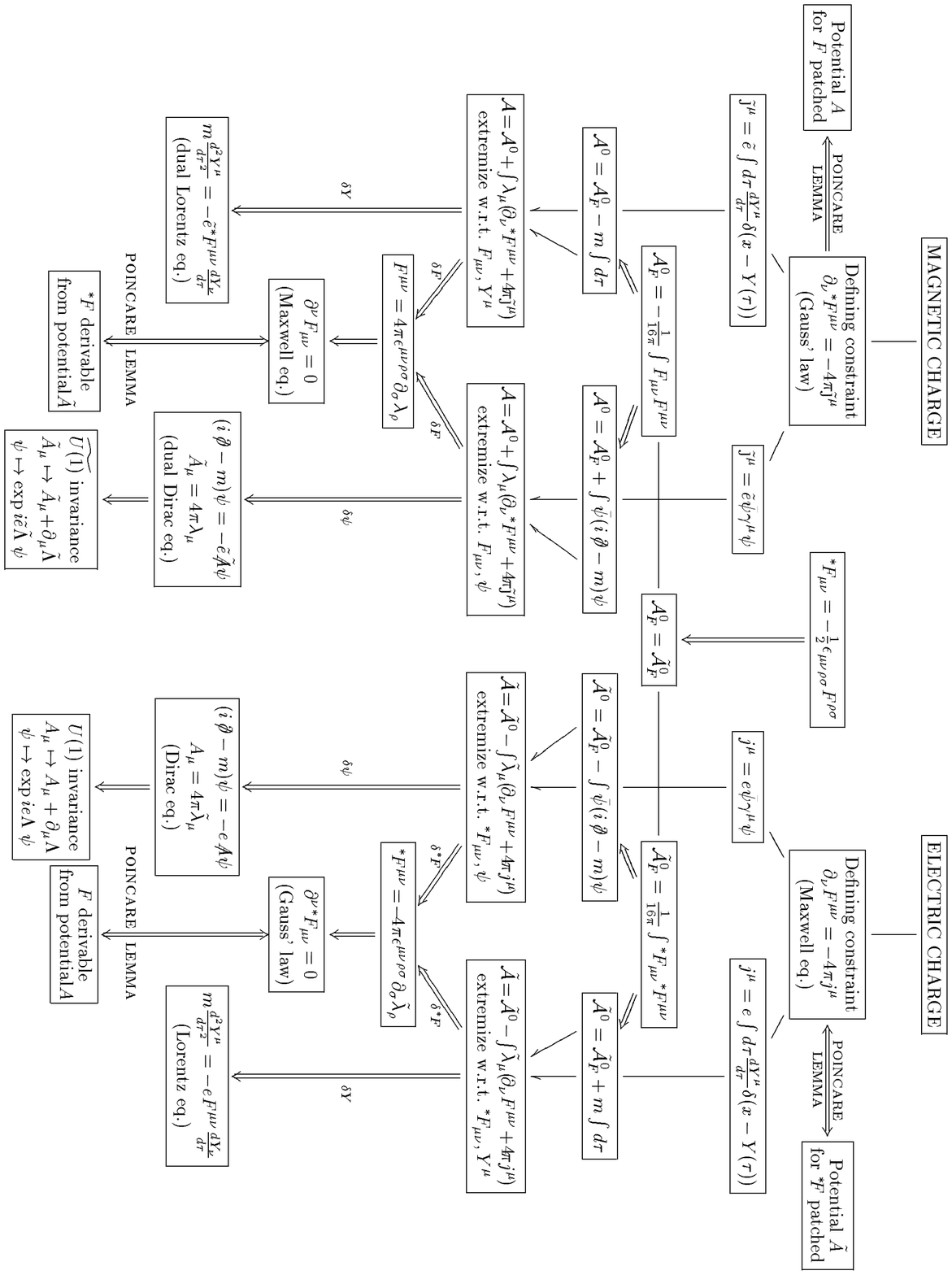}}
    }
\vspace*{-2.2cm}
 \caption[]{Electromagnetism with charges.}
  \label{chart2}
\end{chart}

\clearpage

\section*{Bibliography}
\subsection*{General Gauge theory}
\begin{itemize}
\item IJR Aitchison and AJG Hey, {\em Gauge Theories in Particle Physics}, 
Adam Hilger, 2nd edition, 1989.
\item AP Balachandran et al., {\em Gauge Symmetries and Fibre
Bundles}, Lecture Notes in Physics \# 188, Springer, Berlin, 1983.
\item Chan Hong-Mo and Tsou Sheung Tsun, {\em Elementary Gauge Theory
Concepts}, World Scientific, Singapore, 1993.
\item Tsou Sheung Tsun, Symmetry and symmetry breaking in particle
physics, in {\em Proc.\ 8th EWM General Meeting}, Trieste 1997,
hep-th/9803159. 
\item TT Wu and CN Yang, Concepts of nonintegrable phase factors and
global formulation of gauge fields, {\em Phys.\ Rev.} {\bf D12} (1975)
3845--3857.
\end{itemize}

\subsection*{Mathematics background}
\begin{itemize}
\item N.\ Steenrod, {\em The topology of fibre bundles}, Princeton
University Press, Princeton, 1974.
\item S.\ Kobayashi and K.\ Nomizu, {\em Foundations of Differential
Geometry}, Vol.\ I, Interscience, New York, 1963.
\end{itemize}

\subsection*{Further interests}
Those who are interested in our recent work on duality may like to
read:
\begin{itemize}
\item Chan Hong-Mo and Tsou Sheung Tsun, Nonabelian Generalization of 
  Electric--Magnetic Duality---A
  Brief Review, hep-th/9904102, RAL-TR-1999-014, invited review paper, 
  {\em International J.\ Mod.\ Phys.} {\bf A14} (1999) 2139--2172.
\item Chan Hong-Mo and Tsou Sheung Tsun, The Dualized Standard Model 
  and its Applications---an Interim
  Report, hep-ph/9904406, RAL-TR-1999-015, invited review paper, 
  {\em International J.\ Mod.\ Phys.} {\bf A14} (1999) 2173--2203.
\end{itemize}

\end{document}

%% file: preamble.tex
\input{amssymb.sty}

\def\bbc{{\mathbb C}}

\def\bbr{{\mathbb R}}

\def\bbz{{\mathbb Z}}

\def\Re{\mathop{\rm Re}\nolimits}
\def\Im{\mathop{\rm Im}\nolimits}

\def\tr{\mathop{\rm tr}\nolimits}
\def\Tr{\mathop{\rm Tr}\nolimits}
\def\frac#1#2{{{#1}\over{#2}}}
\def\tfrac#1#2{{\textstyle{{#1}\over{#2}}}}

\def\half{\tfrac{1}{2}}
\def\third{\tfrac{1}{3}}
\def\quart{\tfrac{1}{4}}

\newcommand{\amux}{\mbox{$A_\mu (x)$}}
\newcommand{\fdown}{{\mbox{$F_{\mu\nu}$}}}
\newcommand{\fup}{{\mbox{$F^{\mu\nu}$}}}

\newcommand{\fstarup}{{\mbox{$\mbox{}^*\!F^{\mu\nu}$}}}

\newcommand{\epsup}{{\mbox{$\epsilon^{\mu\nu\rho\sigma}$}}}

\newcommand{\lineint}[3]{
\rlap{$\displaystyle \mathop{\phantom{\int}}\limits_{#3}$}
\int\nolimits_{#1}^{#2}
}

%% file: bohmaharo.pstex_t
\begin{picture}(0,0)%
\epsfig{file=bohmaharo.pstex}%
\end{picture}%
\setlength{\unitlength}{3158sp}%
\begingroup\makeatletter\ifx\SetFigFont\undefined%
\gdef\SetFigFont#1#2#3#4#5{%
  \reset@font\fontsize{#1}{#2pt}%
  \fontfamily{#3}\fontseries{#4}\fontshape{#5}%
  \selectfont}%
\fi\endgroup%
\begin{picture}(6705,2082)(1948,-3391)
\put(2788,-2065){\makebox(0,0)[lb]{\smash{\SetFigFont{7}{8.4}{\rmdefault}{\mddefault}{\updefault}\special{ps: gsave 0 0 0 setrgbcolor}$\Gamma_1$\special{ps: grestore}}}}
\put(2788,-2582){\makebox(0,0)[lb]{\smash{\SetFigFont{7}{8.4}{\rmdefault}{\mddefault}{\updefault}\special{ps: gsave 0 0 0 setrgbcolor}$\Gamma_2$\special{ps: grestore}}}}
\put(5278,-3035){\makebox(0,0)[lb]{\smash{\SetFigFont{7}{8.4}{\rmdefault}{\mddefault}{\updefault}\special{ps: gsave 0 0 0 setrgbcolor}$A_\mu \ne 0$\special{ps: grestore}}}}
\put(5246,-2873){\makebox(0,0)[lb]{\smash{\SetFigFont{7}{8.4}{\rmdefault}{\mddefault}{\updefault}\special{ps: gsave 0 0 0 setrgbcolor}$F_{\mu\nu} = 0$\special{ps: grestore}}}}
\put(5491,-2266){\makebox(0,0)[lb]{\smash{\SetFigFont{7}{8.4}{\rmdefault}{\mddefault}{\updefault}\special{ps: gsave 0 0 0 setrgbcolor}\bf H\special{ps: grestore}}}}
\end{picture}

%% file: loopderiv.pstex_t
\begin{picture}(0,0)%
\epsfig{file=loopderiv.pstex}%
\end{picture}%
\setlength{\unitlength}{2368sp}%
\begingroup\makeatletter\ifx\SetFigFont\undefined%
\gdef\SetFigFont#1#2#3#4#5{%
  \reset@font\fontsize{#1}{#2pt}%
  \fontfamily{#3}\fontseries{#4}\fontshape{#5}%
  \selectfont}%
\fi\endgroup%
\begin{picture}(3497,3590)(3434,-4156)
\put(5041,-4156){\makebox(0,0)[lb]{\smash{\SetFigFont{7}{8.4}{\rmdefault}{\mddefault}{\updefault}\special{ps: gsave 0 0 0 setrgbcolor}$P_0$\special{ps: grestore}}}}
\put(5176,-961){\makebox(0,0)[lb]{\smash{\SetFigFont{7}{8.4}{\rmdefault}{\mddefault}{\updefault}\special{ps: gsave 0 0 0 setrgbcolor}$s$\special{ps: grestore}}}}
\put(6751,-3031){\makebox(0,0)[lb]{\smash{\SetFigFont{7}{8.4}{\rmdefault}{\mddefault}{\updefault}\special{ps: gsave 0 0 0 setrgbcolor}$\Phi^{-1}(C)$\special{ps: grestore}}}}
\put(6931,-1006){\makebox(0,0)[lb]{\smash{\SetFigFont{7}{8.4}{\rmdefault}{\mddefault}{\updefault}\special{ps: gsave 0 0 0 setrgbcolor}$\Phi(C+\delta C)$\special{ps: grestore}}}}
\end{picture}

%% file: holoncurv.pstex_t
\begin{picture}(0,0)%
\epsfig{file=holoncurv.pstex}%
\end{picture}%
\setlength{\unitlength}{3158sp}%
\begingroup\makeatletter\ifx\SetFigFont\undefined%
\gdef\SetFigFont#1#2#3#4#5{%
  \reset@font\fontsize{#1}{#2pt}%
  \fontfamily{#3}\fontseries{#4}\fontshape{#5}%
  \selectfont}%
\fi\endgroup%
\begin{picture}(4995,3926)(1441,-4274)
\put(4321,-2986){\makebox(0,0)[lb]{\smash{\SetFigFont{10}{12.0}{\rmdefault}{\mddefault}{\updefault}\special{ps: gsave 0 0 0 setrgbcolor}$\Gamma_\Sigma$\special{ps: grestore}}}}
\put(2881,-2941){\makebox(0,0)[lb]{\smash{\SetFigFont{10}{12.0}{\rmdefault}{\mddefault}{\updefault}\special{ps: gsave 0 0 0 setrgbcolor}$t=0$\special{ps: grestore}}}}
\put(1441,-2221){\makebox(0,0)[lb]{\smash{\SetFigFont{10}{12.0}{\rmdefault}{\mddefault}{\updefault}\special{ps: gsave 0 0 0 setrgbcolor}$t=t_e$\special{ps: grestore}}}}
\put(6436,-2221){\makebox(0,0)[lb]{\smash{\SetFigFont{10}{12.0}{\rmdefault}{\mddefault}{\updefault}\special{ps: gsave 0 0 0 setrgbcolor}$t=t_e$\special{ps: grestore}}}}
\end{picture}